\documentclass[journal]{IEEEtran}

\usepackage{graphicx}
\usepackage{epstopdf}
\usepackage{subfigure}
\usepackage{amsmath}
\usepackage{amsthm}
\usepackage{amsfonts}
\usepackage{amssymb}
\usepackage{latexsym}
\usepackage{moreverb}
\usepackage{booktabs}
\usepackage{times}
\usepackage{indentfirst}
\usepackage{algorithmic}
\usepackage{algorithm}
\usepackage[table]{xcolor}  
\usepackage{multirow}       
\usepackage{subfigure}

\graphicspath{{eps/}}

\begin{document}
\title{Adaptive Dispatching of Tasks in the Cloud}

\author{Lan Wang and Erol Gelenbe, {\small \em FIEEE}\\
Intelligent Systems and Networks Group\\
Department of Electrical and Electronic Engineering\\
Imperial College London\\
Email: \{lan.wang12,e.gelenbe\}@imperial.ac.uk}

\maketitle

\begin{abstract}
The increasingly wide application of Cloud Computing enables the consolidation of tens of thousands of applications in shared infrastructures. Thus, meeting the QoS requirements of so many
diverse applications in such shared resource environments has become a real challenge, especially since the characteristics and workload of applications differ widely and may change over time. This paper presents
an experimental system that can exploit a variety of online QoS aware adaptive task allocation schemes, and three such schemes are designed and compared.
These are a measurement driven algorithm that uses reinforcement learning, secondly a ``sensible'' allocation algorithm that assigns jobs to sub-systems that are observed to provide a
lower response time, and then an algorithm that splits the job arrival stream into sub-streams 
at rates computed from the hosts' processing capabilities. All of these schemes are compared via measurements
among themselves and with a simple round-robin scheduler, on two experimental test-beds with homogenous and heterogenous hosts having
different processing capacities. 
\end{abstract}


\section{Introduction}
Cloud computing enables elasticity and scalability of computing resources such as networks, servers, storage, applications, and services, which constitute a shared pool, providing on-demand services at the level of infrastructure, platform and software~\cite{mell2009}. This makes it realistic to deliver computing services in a manner similar to utilities such as water and electricity where service providers take the responsibility of constructing IT infrastructure and end-users make use of the services through the Internet in a pay-as-you-go manner. This convenient and cost-effective way of access to services boosts the application of cloud computing, which spans many domains including scientific, health care, government, banking, social networks, and commerce~\cite{buyya2013}. 

An increasing number of applications from the general public or enterprise users are running in the Cloud, generating a diverse set of workloads in terms of resource demands, performance requirements and task execution~\cite{Delimitrou2013}. For example, multi-tier web applications composed of several components which are commonly deployed on different nodes~\cite{Padala2007}, impose varied stress on the respective node, and create interactions across components. Energy consumption remains a major issue \cite{berl2010energy} that can be mitigated through judicious energy-aware scheduling \cite{GelenbeL13,thai13}. 
Jobs being executed in a Cloud environment may be of very different types, such as Web requests that usually demand fast response and produce loads that vary significantly over time~\cite{Jianfeng2013}. 
On the other hand, scientific applications are computation intensive, 
though during execution  they may undergo several phases with varied workload profiles~\cite{Delimitrou2013,Iosup2011}. MapReduce jobs are composed of different tasks of various sizes and resource requirements~\cite{Jianfeng2013}. Furthermore, the nature of cloud computing which enables highly heterogeneous workloads to be served on top of a shared IT infrastructure leads to inevitable interference between co-located workloads~\cite{Zhuravlev}. On the other hand, end users not only rely on the computation resources provisioned by the cloud, but also require assurance of the quality and reliability of the execution of the jobs that they submit. Therefore, the cloud service provider must also dispatch incoming jobs to servers with consideration for the quality of service and cost that it offers within a diverse and complex workload environment.

\subsection{Prior Work}
\label{survey}

Extensive research on this challenging problem has proposed several job scheduling approaches. Static algorithms \cite{tantawi1985optimal,kim1992algorithm,kameda2011optimal} are simple without excessive overhead, but they are only suitable for stable environments, and cannot adapt to changes in a Cloud. Dynamic algorithms ~\cite{wolff2001dynamic,rimal2009taxonomy,zhang2010load,tian2011dynamic} take into consideration different application characteristics and workload profiles both prior to, and during, run-time. They may be quite complex for heterogeneous environments and adapt to dynamic environments, but the resulting computation overhead may cause performance degradation when implemented in a real system. Thus, many of them only evaluated through simulations~\cite{Xiaomin2011} rather than in practical experiments, while some have been tested in a real computer environment with low job arrival rates~\cite{Delimitrou2013}.

Much work on task assignment in the Cloud is based on a detailed representation of tasks to be executed, but a rather simplistic representation of the hosts or processing sub-systems
leading to an evaluation based on simulation experiments rather than measurements on a real system. In \cite{Topcuouglu2002} an application composed of many tasks is represented by a directed acyclic graph (DAG) where tasks, intertask dependency, computation cost, and intertask communication cost are represented; two performance-effective and low-complexity algorithms rank the tasks to assign them to a processor in a heterogeneous environment. Related work is presented in ~\cite{Sih1993,Kwok1996}, while optimization algorithms based on genetic algorithms~\cite{Hou1994}, ant colony optimization (ACO)~\cite{WeiNeng2009}, Particle Swarm Optimization~\cite{Pandey2010}, Random Neural Network optimization~\cite{DBLP:journals/neco/GelenbeF99,Gelenbe2010_nearOptAssign}, and auction-based mechanisms~\cite{Zaman2011} have
also been studied in this context, with potential applications to workload scheduling in the Cloud  \cite{Lin2011}.
	In ~\cite{Moreno2014}, workload models which reflect the diversity of users and tasks in a Cloud production environment are obtained  from a large number of tasks and users over a one month 
	period, and exploited for evaluation in a simulated CloudSim framework.

Other work has used experiments on real test-beds rather than simulations, as in ~\cite{Jianfeng2013} where the characteristics of the typical heterogeneous workloads: parallel batch jobs, web servers, search engines, and MapReduce jobs, results in resource provisioning in a manner that reduces costs for the Cloud itself. Another cost-effective resource provisioning system dedicated to MapReduce jobs ~\cite{Palanisamy2014} uses global resource optimization.
Hardware platform heterogeneity and co-scheduled workload interference are highlighted in ~\cite{Delimitrou2013}, where robust analytical methods and collaborative filtering techniques are use to classify incoming workloads in terms of heterogeneity and interference before being greedily scheduled in a manner that achieves interference minimization and server utilization maximization. The system is evaluated with a wide range of workload scenarios on both a small scale computer cluster and a large-scale cloud environment applying Amazon EC2 to show its scalability and low computation overhead. However, the arrival rate of incoming workload is low and thus the system performance under saturation state is not examined. Furthermore, the contention for processor cache, memory controller and memory bus incurred by collocated workloads are studied in \cite{Zhuravlev2010}.

Early research that consider the important role of servers in delivering QoS in the Internet can be found  in ~\cite{Bhatti1999}, where an architecture is proposed which provides web request classification, admission control, and scheduling with several priority policies to support distinct QoS requirements for different classes of users for multi-tier web applications. However, the scheduling approach is static and in ~\cite{Padala2007}, an adaptive feed-back driven resource control system is developed to dynamically provision resource sharing for multi-tier applications in order to achieve both high resource utilization and application-level QoS. A two-tiered on-demand resource allocation mechanism is presented in ~\cite{Ying2013} with local allocation within a server and global allocation based on each local one, so as to achieve better resource utilization and dynamically adjust according to time-varying capacity demands. Energy consumption in computation, data storage and communications is also a challenge in the cloud~\cite{berl2010energy}. A model for server performance and power consumption is derived in ~\cite{douratsos} with the potential to predict power usage in terms of workload intensity. In ~\cite{gelenbe2012trade,thai13}, the authors examine the selection of system load that provide the best trade-off between energy consumption and QoS. A heterogeneity-aware dynamic capacity provisioning scheme for cloud data centers is proposed in ~\cite{Zhang2014}, which classifies workloads based on the heterogeneity of both workload and machine hardware and dynamically adjusts the number of machines so as to optimize overall energy consumption and scheduling delay.

\subsection{Overview of our Approach}

The present paper uses experiments to investigate adaptive dynamic allocation algorithms that take decisions based on up-to-date measurements,
and make fast online decisions to attempt achieve desirable QoS levels~\cite{dillon2010cloud}. The software that we have designed to this effect is a practical system 
implemented as a Linux kernel module which can be easily installed and loaded on any PC with the Linux OS. Its design is inspired by Cognitive Packet Network \cite{SAN04} which is a QoS-driven adaptive routing protocol 
that select paths in a network to provide the best possible QoS for the network's traffic based on online measurement. 

In the approach we propose, we embed measurement agents into each host in a cloud to observe the state system. These observations are then collected by ``smart packets'' (SPs) that are sent at regular intervals 
into the system in a manner which favours the search of those sub-systems which are of the greatest
interest because they may be used more frequently or because they could provide better performance. The SPs then come back to the controller which uses a dynamic algorithm based either on the Random Neural Network (RNN)~\cite{SAN04,DBLP:journals/neco/GelenbeT08}, or on a form of online greedy adaptation called ``sensible routing'' ~\cite{gelenbe2003sensible} that selects probabilistically
the host whose measured QoS is the best.  We also study a task allocation scheme that splits the incoming jobs arrival into streams towards the different hosts, at fixed arrival rates 
chosen so as to take the best advantage of the hosts' relative processing ability. We have conducted experiments with the RNN-based scheme, the sensible routing scheme, and the fixed rate scheme under varied job arrival rate via experiments on a real computer cluster test-bed, and compared them with other static algorithms such as Round Robin and an allocation scheme that distributes the jobs equally between hosts. To further our investigation, we have set up two different cluster test-beds: one composed of hosts with relative uniform processing ability, and the other one with an increasing processing capacity difference between hosts. The resulting experimental results are carefully analyzed and reported.

The remainder of the paper is organized as follows. We detail the novel task allocation platform that we propose in Section \ref{task}, where the dynamic algorithms are introduced. Section \ref{math} proposes a mathematical model for our system, which leads to the design of a fixed arrival rate based allocation scheme. Experimental results are presented in Section \ref{experimentalresults} to  compare the performance obtained with all of the above allocation schemes. Section \ref{conclusions} draws our main conclusions and discusses directions
for future research. 

\section{Task Allocation Platform and Test-Bed} \label{task}
\label{research}



In this section we propose a task allocation platform (TAP) where online monitoring and measurement are constantly carried out in order to keep track of the state of the cloud system, including the current resource utilisation (CPU, memory, and I/O), the system load, the application-level QoS requirements, such as job response time and bandwidth, as well as energy consumption, and possibly also (in future versions of the system) system security and economic cost. With knowledge learned from these observations,  the system employs the QoS driven task allocation algorithms that we have designed, to make online decisions that can achieve the best possible QoS as specified by the tasks' owners, while adapting to varying conditions over time. 

\begin{figure}[ht]
     \centering
      \includegraphics[scale=0.45,clip=true, trim=10 400 90 150]{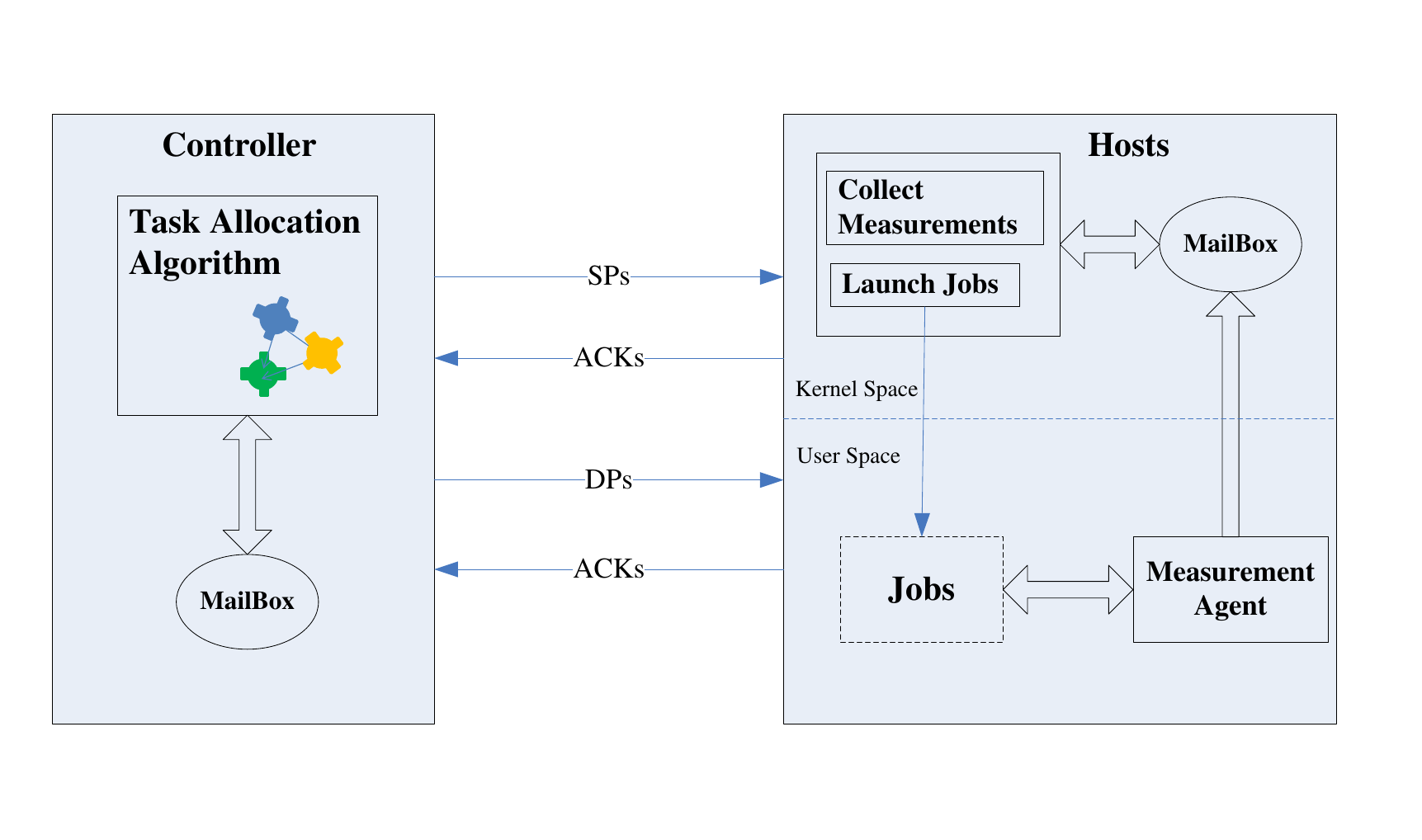} 
     \caption{System Architecture of the Task Allocation Platform (TAP)}
     \label{fig:platform}
    \end{figure}

Figure~\ref{fig:platform} shows the building blocks of our platform. The controller, which is the intellectual center of the system, accommodates the online task allocation algorithms, which usually work along with a learning algorithm, with the potential to adaptively optimize the use of the cloud infrastructure. Our platform penetrates into the cloud infrastructure by deploying measurement agents: these agents conduct online observations that are relevant to the QoS requirements of end users, and send back the measurements to the controller. Using ideas from ``Cognitive Packet Network'' routing in packet networks \cite{TAI,network1}, three types of packets are used for communications between the components within the platform: smart packets (SPs) for discovery and measurement, dumb packets (DPs) for carrying job requests or jobs, and acknowledgement packets (ACKs) that carry back the information that has been discovered by SPs and experienced by allocated jobs. In this section, we present in detail the mechanisms that are implemented  in the platform and the algorithms that are used.

We conduct our experiments on a real a test-bed cluster composed of four nodes. One node is dedicated to the TAP, and the other three nodes are used as hosts running jobs, as shown in Figure~\ref{fig:testbed},
with each having a different processing power so that we may observe significant execution time differences for a given job running in each of the clusters. 
TAP takes decisions based on online measurements which collected by SPs. Even when there are no incoming jobs, the system maintains awareness of the state of the cloud by sending SPs periodically. End users are allowed to declare the QoS requirements related to the jobs they are planning to submit, which is then translated into one or more QoS metrics which constitute a function called the ``goal function'' in our system. In this way, the QoS requirements are transformed into a goal function to be minimized, e.g. the minimization of the job response time. The goal function determines which system parameters need to be measured and how optimized task allocation is to be carried out. 

SPs are first sent at random to the various hosts in order to obtain some initial information and inform the measurement agents in the hosts to activate the requested measurement. The task allocation algorithm in TAP learns from the information carried back by the ACKs and makes adaptively optimized decisions which are used to direct the subsequent SPs. Thus, the SPs collect online measurements in an efficient manner and pay more attention to the part of the cloud where better QoS can be offered, visiting the worse performing parts less frequently. 

The incoming jobs or job requests are encapsulated into the DPs, and exploit the decisions explored by SPs to select the host/cloud sub-system that will to execute the job. Once a job (request) arrives at a host in the cloud, its monitoring is started by the measurement agent which records the trace of the job execution until it is completed and deposits the records into a mailbox which is located in the kernel memory of the host. When a SP arrives at this host, it collects the measurements in the mailbox and generates an ACK which carries the measurements, and travels back to the controller where the measurement data is extracted and used for subsequent decisions of the task allocation algorithm. As soon as a job completes its execution, the agent also produces an ACK heading back to the controller with all the recorded data, such as the job arrival time at the cloud, the time at which the job started running and the time at which the job execution completed. When the ACK of the DP reaches the controller, the job response time at the controller is estimated by taking the difference between the current arrival time at the node and the time at which the corresponding job arrives at the controller which is used by the algorithm when the job response time is required to be minimized. 

TAP may use different schemes to make decisions regarding task allocation, and in the sequel we will describe two randomized schemes in Section \ref{random}, as well as a scheme based on Reinforcement Learning \cite{Sutton} which uses the random neural network model \ref{RNN} as the adaptive critic for the goal or cost function to be minimized.

\subsection{TAP's Randomized Task Allocation Schemes} \label{random}

By a {\em randomized task allocation scheme} for TAP, we mean that when a job arrives at TAP from some user or source outside the Cloud system, TAP decides to allocate it to some host $i$ among the $N$
available hosts, with probability $p_i$ so
that {\em at decision time when the task must be allocated}:
\begin{itemize}
\item TAP first calculates $p_i$ for each of the hosts $i$,
\item Then TAP uses these probabilities to actually select the host that will receive the task.
\end{itemize}
Randomized schemes have the advantage that a host which is being preferred because it is providing better service is {\em not} being overloaded by repeated allocation since the QoS it offers is
only used probabilistically to make a task allocation.

To this effect, TAP uses two distinct schemes to calculate $p_i$, Sensible Routing, and Model Based Allocation.
\subsubsection{Sensible Routing}
The sensible decision algorithm had been proposed in~\cite{gelenbe2003sensible} as an adaptive routing algorithm which applies randomized routing policies based on the expected QoS so as to improve QoS. We use the sensible decision algorithm in our task allocation system where allocation policies are described as probabilistic choices among all available hosts. QoS metrics are defined as non-negative random variables. A QoS metric is viewed as sensitive when the value for the QoS metric corresponding to a host increases as the probability of dispatching jobs to that host increases. Job response time and job execution time are examples of sensitive metrics.
 
 In the Sensible Routing approach that we propose for TAP,  we use a weighted average $G_i$ of the goal function that we wish to {\em minimize}. $G_i$ is  estimated for
for each of the hosts $i$, and updated each time $t$ that TAP receives a measurement that can be used to update the goal function. If the measured total job response time $G_i^t$
is received at the TAP regarding host $i$, then the following the expression is used to update TAP's estimate of $G_i$:
\begin{equation}
G^{n}_i=(1-\alpha) G^{n-1}_i + \alpha G_i^t,
\end{equation}
where the parameter $0\leq \alpha\leq 1$ is used to vary the weight given to the most recent measurement as compared to past values, and $n$ denotes the value of the goal function obtained
after the $n$-th update. The probability $p_i^s$ that may then be used to allocate a task to a host will be:
\begin{equation}
p_i^s = \frac{\frac{1}{G^n_i}}{\sum_{j=1}^N\frac{1}{G^n_j}}. 
\end{equation}
Of course, when TAP needs to allocate a task, it will use the most recent value of $p_i^s$ which is available.

\subsubsection{Model Based Task Allocation} \label{math}

Model Based Allocation uses a mathematical model to predict the estimated performance at a host in order to make a randomized task allocation. This has been used in earlier work concerning
task allocation schemes that help reduce the overall energy consumed in a system \cite{thai13}. In this approach, if $W_i(\lambda,p_i)$ is the relevant QoS metric obtained
for host $i$ by allocating a randomized fraction $p_i$ of jobs to host $i$ when the overall arrival rate of jobs to TAP is $\lambda$, then the allocation probabilities $p_1,~...~,p_N$ are chosen so as to minimize the
overall average QoS metric:
\begin{equation}
W = \sum_{i=1}^N p_iW_i(\lambda,p_i).
\end{equation}
At first glance, since each host $i$ is a multiple-core machine with $C_i$ cores, a simple mathematical model that can be used to compute, say the QoS metric
``response time'' $W_i(\lambda,p_i)$ that host $i$ provides, assuming that there
are no main memory limitations and no interference among processors (for instance for memory or disk access), is the $M/M/C_i$ queueing model \cite{Gelenbe-Mitrani}, i.e. with Poisson arrivals, exponential service times, and $C_i$ servers. Of course, both the
Poisson arrival and the exponential service time assumptions are simplifications of reality, and more detailed and precise models are also possible for instance using
diffusion approximations \cite{Diffusion} but would require greater computational effort and more measurement data.

\begin{figure}[ht]
     \centering
      \includegraphics[scale=0.5,clip=true, trim=100 250 100 300,width=6.5cm,height=4.3cm]{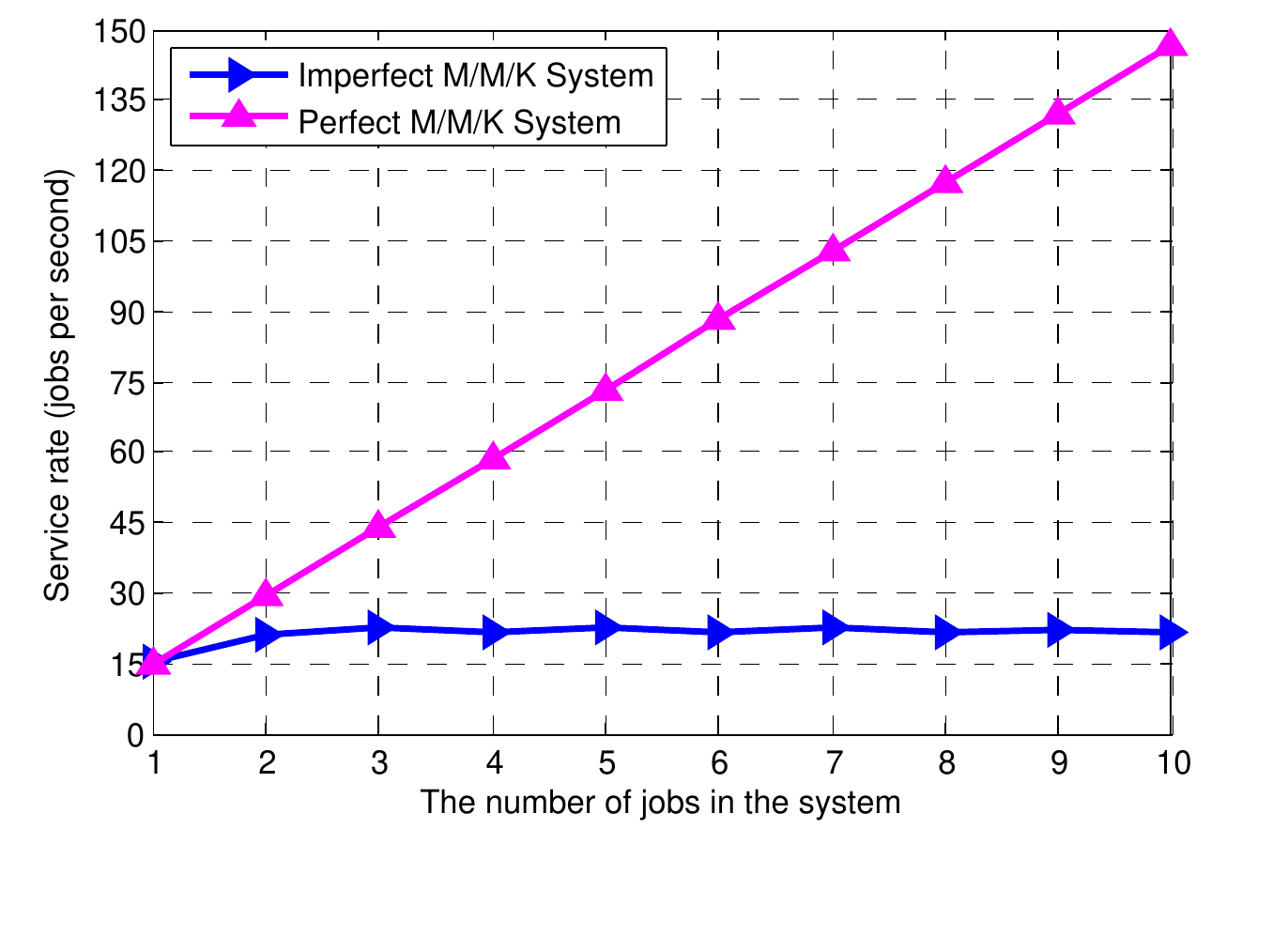} 
     \caption{The idea service rate provided by the perfect multiple core system (red), compared to the measured job completion rate on Host 1 (blue),
  plotted against the number of jobs running simultaneously on the host (x-axis).}
     \label{fig:service_rate}
    \end{figure}

However, a set of simple experiments we have conducted show that the $M/M/K$ model for each host would not correspond to reality. Indeed, in Figure \ref{fig:service_rate} we report the {\em measured}
completion rate of jobs on a host (y-axis) relative to the execution time for a single job running by itself, as a function of the number of simultaneously running
jobs (x-axis). These measurements were conducted on a single host (Host 1), and for a single job running on the system, the average
job processing time was $64.1$ms.

\begin{figure}[ht]
     \centering
      \includegraphics[scale=0.5,clip=true, trim=100 250 100 300,width=6.5cm,height=4.3cm]{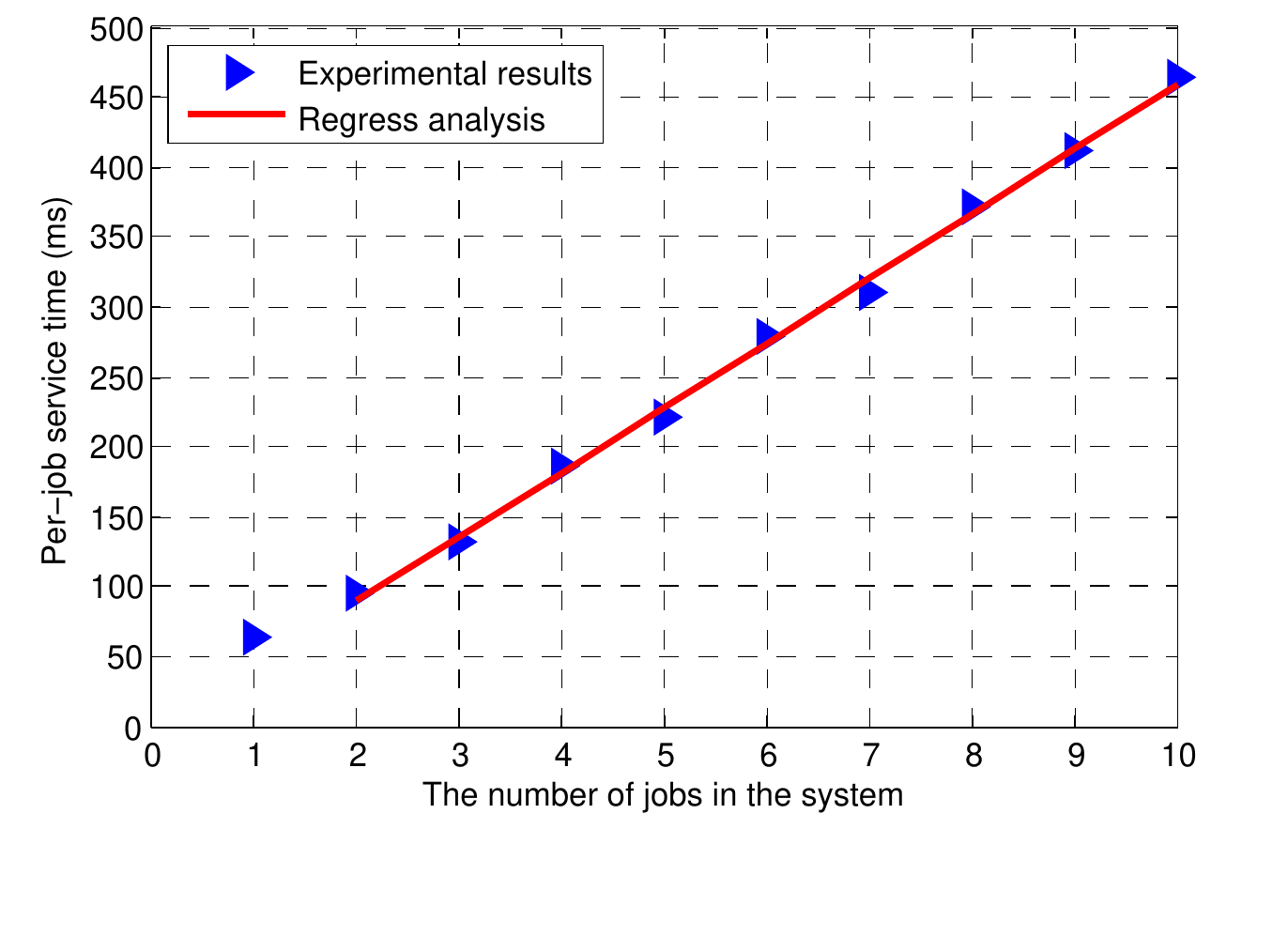} 
     \caption{Measurement of the effective job execution time per job on Host 1, versus the number of simultaneously running jobs 
     on the host (x-axis).}
     \label{fig:service_time_perjob}
    \end{figure}

If this were a perfectly running ideal parallel processing system, we could observe something close to a linear increase in the completion rate of jobs (red dots) when the number of simultaneously running jobs increases, until the
number of cores in the machine $C_1$ have been reached. However the measurements shown in Figure\ref{fig:service_rate} indicate (blue dots)
a significant increase in completion rate as the number of jobs goes from $1$ to $2$, but then the rate remains constant, which reveals that there may be significant interference between jobs due to competition for resources. Indeed, if we call $\gamma(l)$ the average completion rate per job, we observed the following values for $\gamma_i(l)/\gamma_i(1)$ for $l=2,~...~,10$ computed to two decimal digits:
$0.67, 0.48, 0.34,0.29,0.23, 0.20,0.17, 0.15, 0.13$. From this data, a linear regression estimate was then computed for the average execution time $\mu(i)^{-1}$ when there
are $l$ jobs running simultaneously, as shown on Figure \ref{fig:service_time_perjob}, yielding a quasi-linear increase. As a result we can quite accurately
use the estimate $l.\gamma(l)/\gamma(1)\approx 1.386$. 
Based on this measured data, we model the distribution of the number of jobs in a host server $i$ as a random walk on the non-negative integers,
where:
\begin{itemize}
\item $l=0$ represents the empty host (i.e. with zero jobs at the host), 
\item The transition rate from any state $l\geq 0$ to state $l+1$ is the arrival rate of jobs to the host
$\lambda_i$, 
\item The transition rate from state $1$ to state $0$ is the $\mu_i(1)=T_i^{-1}$ where $T_i$ is the average execution time of a job (by itself) on the host,
\item The transition rate from state $l+1$ to state $l$ if $l\geq 1$ is quasi constant given by $\mu_{i0}\equiv (l.\gamma(l)/\gamma(1))\mu_i(1)$,
\item The arrival rate of jobs to Host $i$ is $\lambda_i=p_i^m\lambda$ where $p_i^m$ is the probability with which TAP using the model based algorithm assigns 
jobs to Host $i$, and $\lambda$ is the overall arrival rate of jobs to TAP.
\end{itemize}
The probability that there are $l$ jobs at Host $i$ in steady-state is then:
\begin{eqnarray}
p_i(1)=p_i(0)\frac{\lambda_i}{\mu_i(1)},\nonumber\\
p_i(l)=(\frac{\lambda_i}{\mu_{i0}})^{l-1}p_i(1),~l>1,\nonumber\\
p_i(0)=\frac{1-\frac{\lambda_i}{\mu_{i0}}}{1+\lambda_i\frac{\mu_{i0}-\mu_i(1)}{\mu_{i0}\mu_i(1)}}.\nonumber
\end{eqnarray}
and the resulting average response time for jobs arriving to Host $i$, which by Little's formula \cite{Gelenbe-Mitrani} is 
equal to the average number of the jobs divided by the arrival rate at Host $i$, becomes:
\begin{equation}
W_i^m=\frac{1}{\mu_i(1)}\frac{p_i(0)}{(1-\frac{\lambda_i}{\mu_{i0}})^2},
\end{equation}
and the overall average response times that we wish to minimize, by chosing the
$p_i^m$ for a given $\lambda$ is:
\begin{equation}
W^m=\sum_{i=1}^N\frac{p_i}{\mu_i(1)}\frac{p_i(0)}{(1-\frac{\lambda_i}{\mu_{i0}})^2}.
\label{eq:minavgresptime}
\end{equation}
The appropriate values of the $p_i^m$ for a given system and a given arrival rate $\lambda$ can be then obtained numerically.

\begin{figure}[ht]
     \centering
      \includegraphics[scale=0.5,clip=true, trim=100 250 100 250]{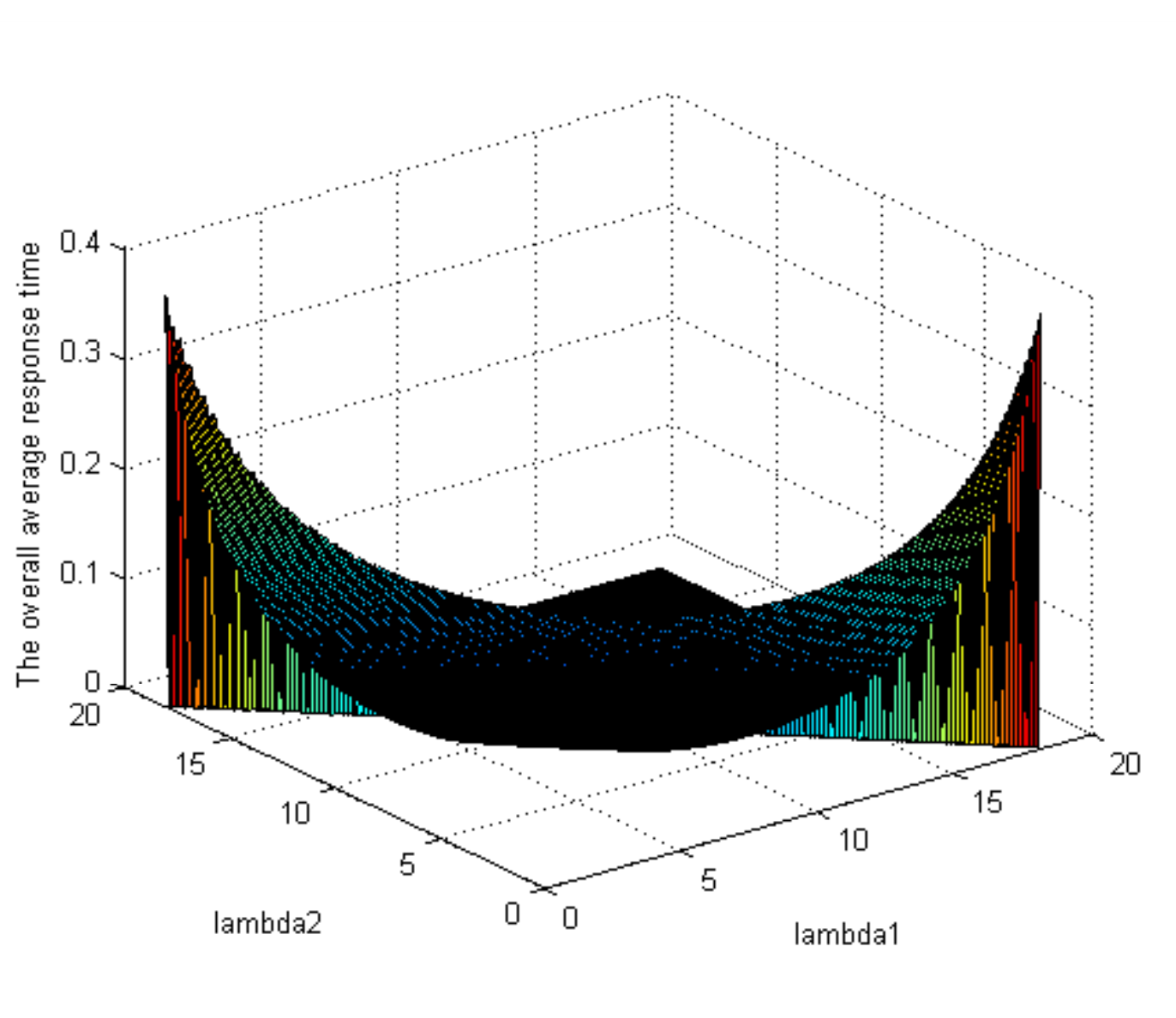} 
     \caption{Variation of the overall average job response time predicted by the infinite server model, with different combinations of $[\lambda_1,\lambda_2,\lambda_3]$, when $\lambda=\lambda_1+\lambda_2+\lambda_3$ is set to 20 jobs per second.}
     \label{fig:fixlambda}
    \end{figure}

To illustrate this approach for the specific service time data regarding the three hosts that we use in our test-bed, 
in Figure~\ref{fig:fixlambda} 
we show the variation of the average job response time with different combinations of $[\lambda_1,\lambda_2,\lambda_3]$,
when  $\lambda=20~jobs/sec$.  

\subsection{Random Neural Network-based Allocation} \label{RNN}

The Random Neural Network (RNN) has been used in static task allocation problems~\cite{Gelenbe2010_nearOptAssign}, as well as for dynamic allocation of traffic to routes in packet networks \cite{network1}. The RNN comprised of $N$ neurons is often used in a ``recurrent'' or fully connected form \cite{RNN}, where each neuron $i$ is characterized by an integer $k_i(\tau)\geq 0$ where $\tau$ represents time, and each neuron is connected to other neurons by both excitatory and inhibitory weights. Furthermore, for the specific application that we are considering in the TAP, each neuron is identified with a particular host, i.e. neuron $i$ is identified with the
decision to assign a task to Host $i$. The theoretical underpinning of the RNN \cite{Gelenbe1990_RNN,Gelenbe01112008} is based on a theorem that states that, at the equilibrium state,
the probabilities:
\begin{equation}
q_i=\lim_{\tau\to\infty} Prob[k_i(\tau)>0],
\end{equation}
are obtained from the expression:
\begin{equation}
\label{q}
	q_i = \frac{\Lambda(i)+\sum_{j=1}^Nq_jw^+(j,i)}{r(i)+\lambda(i)+\sum_{j=1}^Nq_jw^-(j,i)},
	\end{equation}
where the $w^+(j,i)$ and $w^-(j,i)$ are the excitatory and inhibitory weights from neuron $j$ to neuron $i$, and
$\Lambda(i)$ and $\lambda(i)$ are the external flows or inputs of external excitatory and inhibitory signals to neuron $i$,
while $r(i)$ is the firing or total activity rate of neuron $i$:
\begin{equation}
r(i) = \sum_{j=1}^{N}[w^+(i, j) + w^-(i, j)]
\label{ri}
\end{equation}
In the present case, we assume that a distinct RNN is set up within the TAP to cover each distinct goal function. However, these
different RNNs would not have to be created in advance and stored at the TAP indefinitely, but instead created when they are actually needed. Thus we imagine that
we may have a different RNN that is used to decide about allocations that involve minimizing the economic cost of a task allocation (as when the end users are expected to pay a monetary price for the work they receive),
or a different one that deals with minimizing response time, and so on. 

Suppose that the goal function to be minimized is denoted by $G$, such as the response time 
to incoming jobs or the execution time of jobs. Before collecting any measurements in the system we initialize the 
decision system with a parameter set  $q_i=0.5$ for all of the $i$, obtained by setting
$w^+(i,j)=w^-(i,j)=1/2(N-1)$, and set all ``self'' excitation and inhibition rates (from the neuron to itself)
to zero. Thus $r(i)=1$ for all $i$, and $\Lambda(i)=0.25+0.5\lambda(i)$. In particular we can choose $\lambda(i)=0$ so that all $\Lambda(i)=0.25$.

TAP will then use the $q_i,~i=1,~...~,N$ to make allocations so that the task is assigned to the host with the highest value of $q_i$, and in the initial value chosen
any one of the hosts will be chosen with equal probability.
However with successive updates of the weights, this will change so that TAP will select the ``better'' hosts which provide a smaller value of $G$. 

Thus when TAP receives an observation or measurement $G_i^t$ with regard to the goal function that one wishes to {\em minimize}, the RNN weights are updated as follows: 
\begin{itemize}
\item We first update a decision threshold $T_l$ as
\begin{equation} \label{thr}
	T_l= \alpha T_{l-1} + (1-\alpha)G_i^t
	\end{equation}
where $0<\alpha < 1$ is a parameter used to vary the relative importance of ``past history''.
\item Then, if $G_i^t<T_l$ then it is considered that the advice provided by the RNN in the past was successful and
TAP updates the weights as follows:
\begin{eqnarray}
w^+(j,i) &\leftarrow& w^+(j,i) +  G^t_i  \nonumber\\
w^-(j,k) &\leftarrow& w^-(j,k) +  G^t_i/(n-2), \hspace{1mm}  if \hspace{1mm} k \neq i \nonumber
\end{eqnarray}
\item $else~if~G^t_i>T_l$
\begin{eqnarray}
w^+(j,k) &\leftarrow& w^+(j,k) +  G^t_i/(n-2), \hspace{1mm}  if \hspace{1mm} k \neq i \nonumber  \\
w^-(j,i) &\leftarrow& w^-(i,j) + G^t_i  \nonumber
\end{eqnarray}
\end{itemize}
After the weights are updated, the $q_i$ are computed again with new weights.
We note that this algorithm will tend to increase the probability $q_i$ of those neurons which correspond to hosts that yield a better value of
$G_i$, which is why each time TAP assigns a task to a host, it uses the host $i$ that corresponds to the largest $q_i$.

In order to make sure that TAP tries out other alternates and does not miss out on better options, 10\% of the decisions are made at random: thus on average
one out of ten decisions are based on a random (equally likely) choice among all hosts, while 90\% of the decisions are based on the optimization algorithm that we have described.

Note also that this algorithm can be modified to a ``sensible'' version where:
\begin{equation}
p_i^{RNN-S}=\frac{q_i}{\sum_{j=1}^Nq_j}. \label{RNN-S}
\end{equation}

\section{Experiments with the Task Allocation Platform}
\label{experimentalresults}

\begin{figure}[ht]
     \centering
      \includegraphics[scale=0.5,clip=true, trim=40 130 40 380]{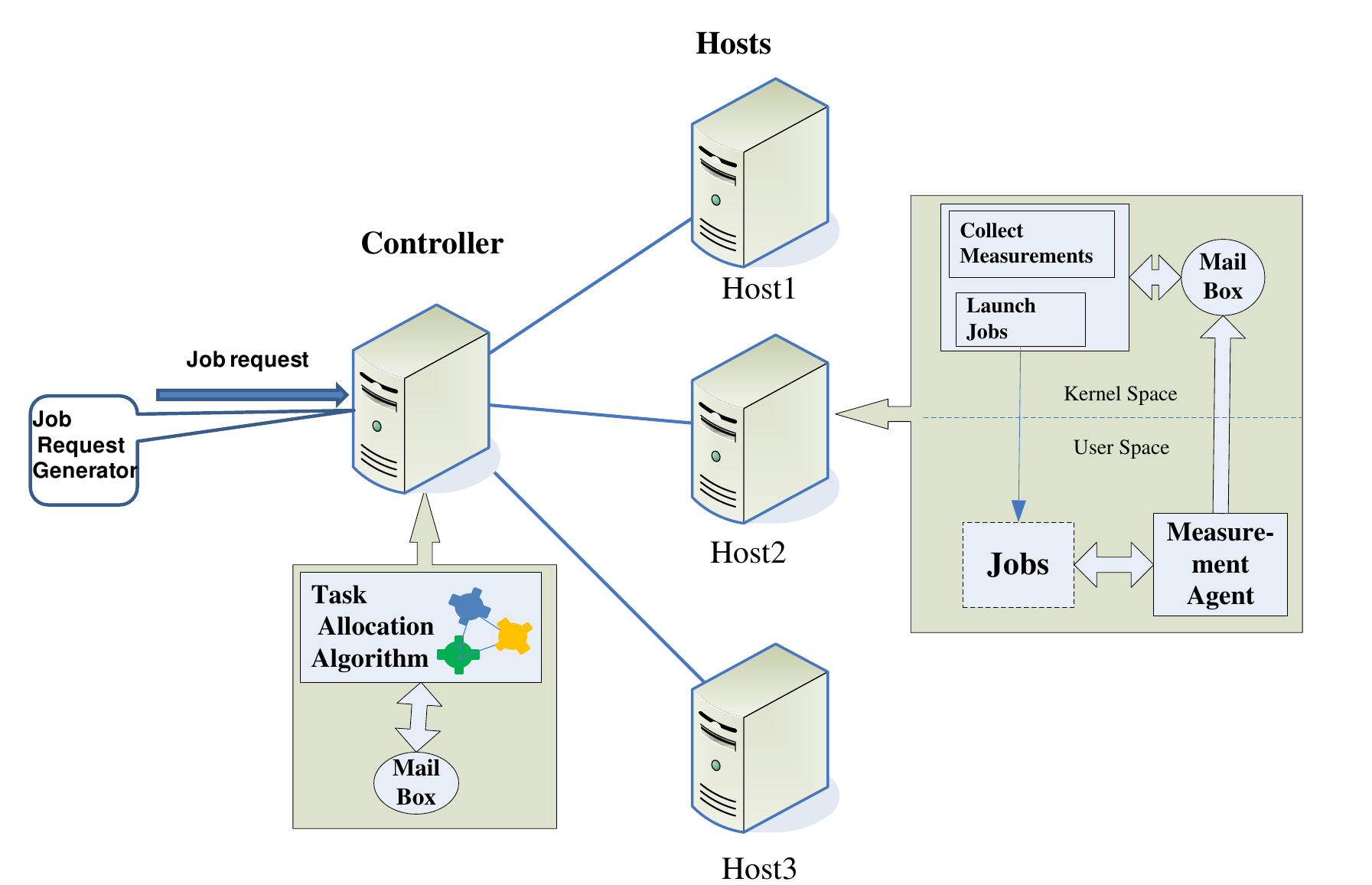} 
     \caption{Task allocation testbed}
     \label{fig:testbed}
    \end{figure}

Our proposed platform TAP is a practical system which can exploit several different task allocation algorithms, such as the three that have been described above, and
it is implemented as a Linux kernel module which can be easily installed and loaded on any PC with Linux OS.
We have implemented TAP for a cluster of three hosts for job execution and a separate host working as the controller, to which the three hosts are connected, as shown in Figure~\ref{fig:testbed}. 

For the purpose of our experiments, a synthetic benchmark is generated, with job profiles indicated  by using the  fields $\{Job~ID, QoS~requirement,Job~Size\}$, which are then packetized into an IP packet and sent to the controller. The job request generator can be configured to send job requests either at a fixed inter-job interval, denoted ``constant rate'' or CR, or following a Poisson process with independent and identically distributed
inter-job arrival intervals with a given rate (denoted EXP).  The controller where TAP is running, decides on job placement based on the measurements carried back by ACKs and deposited in the mailbox. 

The experiments we report in this paper were run with jobs that were defined as a ``prime number generator with an upper bound $B$ on the prime number being generated''. Thus the choice of $B$ allowed us to vary both
the execution time and the memory requirements of the job. We did not actually ``transfer'' the jobs from the task controller to the host, but rather installed the job in advance on the host,
and the allocation decision by TAP just resulted in arrival of a message from TAP to activate the job with specific value of $B$ on that particular host. 
The measurement agent resident on that host then monitored the job execution and recorded its measurements into the mailbox. Both the jobs and the measurement agent run in the user's memory space,
while  the module that receives the SPs and job requests carried by DPs, collects measurements from the mailbox, and generates ACKs with the collected measurements runs in the kernel space of memory as shown in Figure~\ref{fig:testbed}, so that interference between the user program and the system aspects are avoided at least within the memory. 


We set up the six experimental scenarios listed in Table~\ref{tab:experimentscenarios}. The two QoS goals that were considered were (a) the minimization of either the execution time on the host, and (b) the minimization of the response
time at TAP, which includes the message sent to activate the job at a host and the time it takes for an ACK to provide information back to the TAP, i.e. job execution time and job response time at the controller. 

We first  used TAP with the RNN algorithm with Reinforcement Learning (RL) as described above, and TAP with the sensible decision algorithm, and compared their performance. 

The RNN based TAP was experimented with both (a) and (b), whereas the sensible decision based TAP only
used (b) the job response time at the controller.  In addition, according to the analytical model based approach was  with (b) job response time 
computed in terms of the job arrival rate and the system service rate, and then used to determine the optimum values of $\lambda_1$, $\lambda_2$, $\lambda_3$ 
corresponding to the three hosts subject to $\lambda=\lambda_1+\lambda_2+\lambda_3$, with an aim to minimize the overall job response time of the system as in (\ref{eq:minavgresptime}), and
then conducted experiments with job allocation probabilities to the three hosts selected
so as to result in the arrival streams to the three hosts having the rates recommended by the analytical solution. 

We also compared  two static allocation schemes: Round Robin where successive jobs are sent to each host of the cluster in turn, and an equally probable allocation where a job is dispatched to 
each host with equal probability $0.33$.

All these experiments were repeated for a range of average job arrival rates $\lambda$ equal to $1, 2, 4, 8, 12, 16, 20, 25, 30, 40$ jobs per second. Each experiment lasted 5 mins so as to achieve a stable state for each experiment.

	\begin{table}[ht]
	\centering
	\begin{tabular}[scale=0.35]{l|l}
				\hline
				Notation&Description\\
				\hline
				RNNs (RT) with CR & Random Neural Network algorithm \\ &with online measurement of the \\ &job response time at the controller \\ &and constant job arrival rates\\
				\hline
				RNNs (RT) with EXP & Random Neural Network algorithm \\ &with online measurement of the \\ &job response time at the controller \\ &and exponentially \\&distributed job interarrival time\\
				\hline
				RNNs (ET) with CR & Random Neural Network algorithm \\ &with online measurement of the \\ &job execution time and constant \\ &job arrival rates\\
				\hline
				RNNs (ET) with EXP & Random Neural Network algorithm \\ &with online measurement of the \\ &job execution time and exponentially \\ &distributed job interarrival time\\
				\hline
				Sensible Decision with CR & Sensible Decision algorithm \\ &with online measurement of the \\ &job response time at the controller and \\ &constant job arrival rates\\
				\hline
				Sensible Decision with EXP & Sensible Decision algorithm \\ &with online measurement of the \\ &job response time at the controller and \\ &exponentially distributed job \\ &interarrival time\\
				\hline
			
			\end{tabular}  \\
		\caption{Experiment Scenarios}
	\label{tab:experimentscenarios}
\end{table}

		\begin{figure}[ht]
     \centering
      \subfigure[\label{fig:cr_exp_exe}]{\includegraphics[scale=0.5,clip=true, trim=100 250 100 300,width=6.5cm,height=4.2cm]{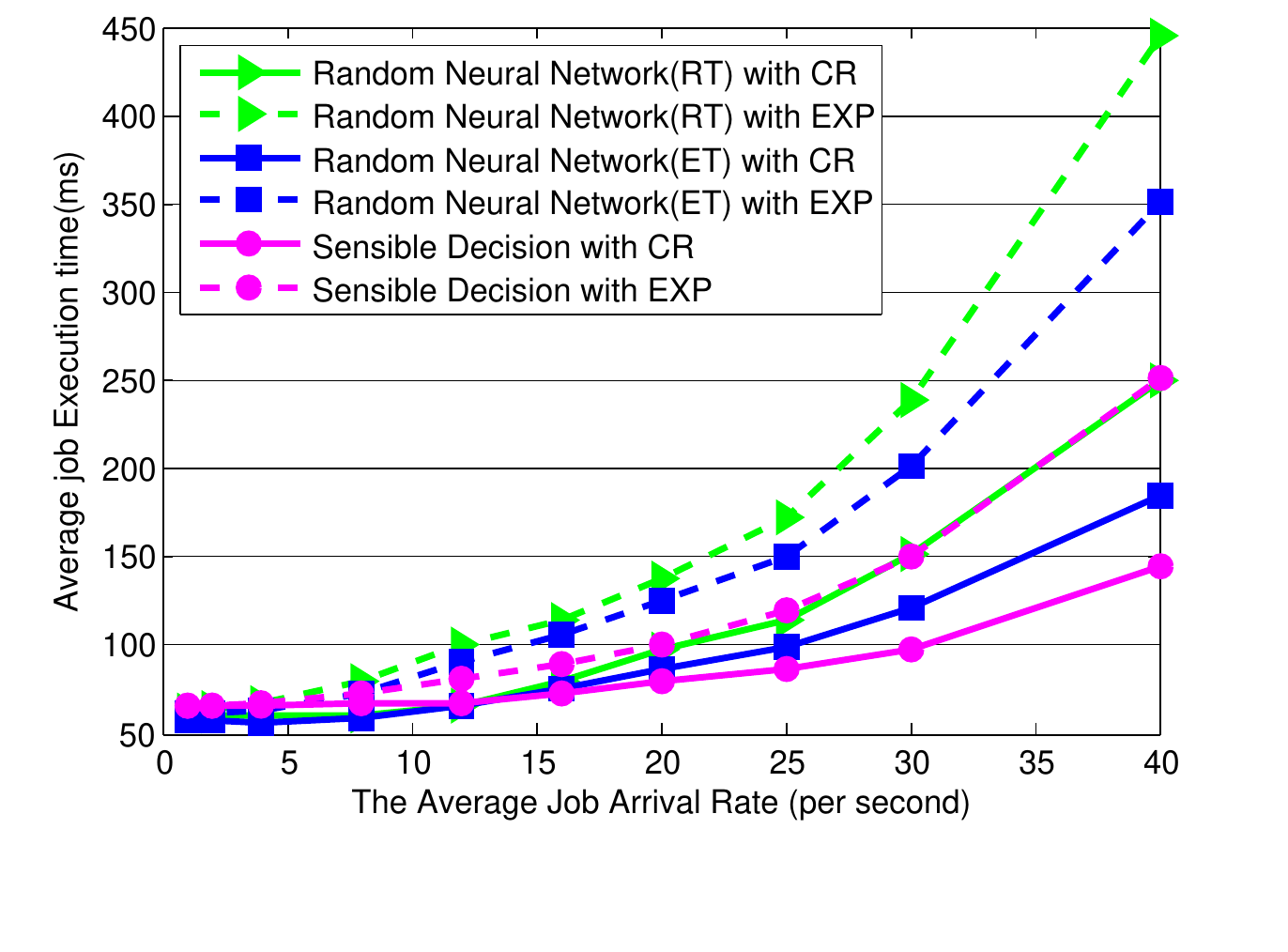}}
      \subfigure[\label{fig:cr_exp_host}]{\includegraphics[scale=0.5,clip=true, trim=100 250 100 300,width=6.5cm,height=4.2cm]{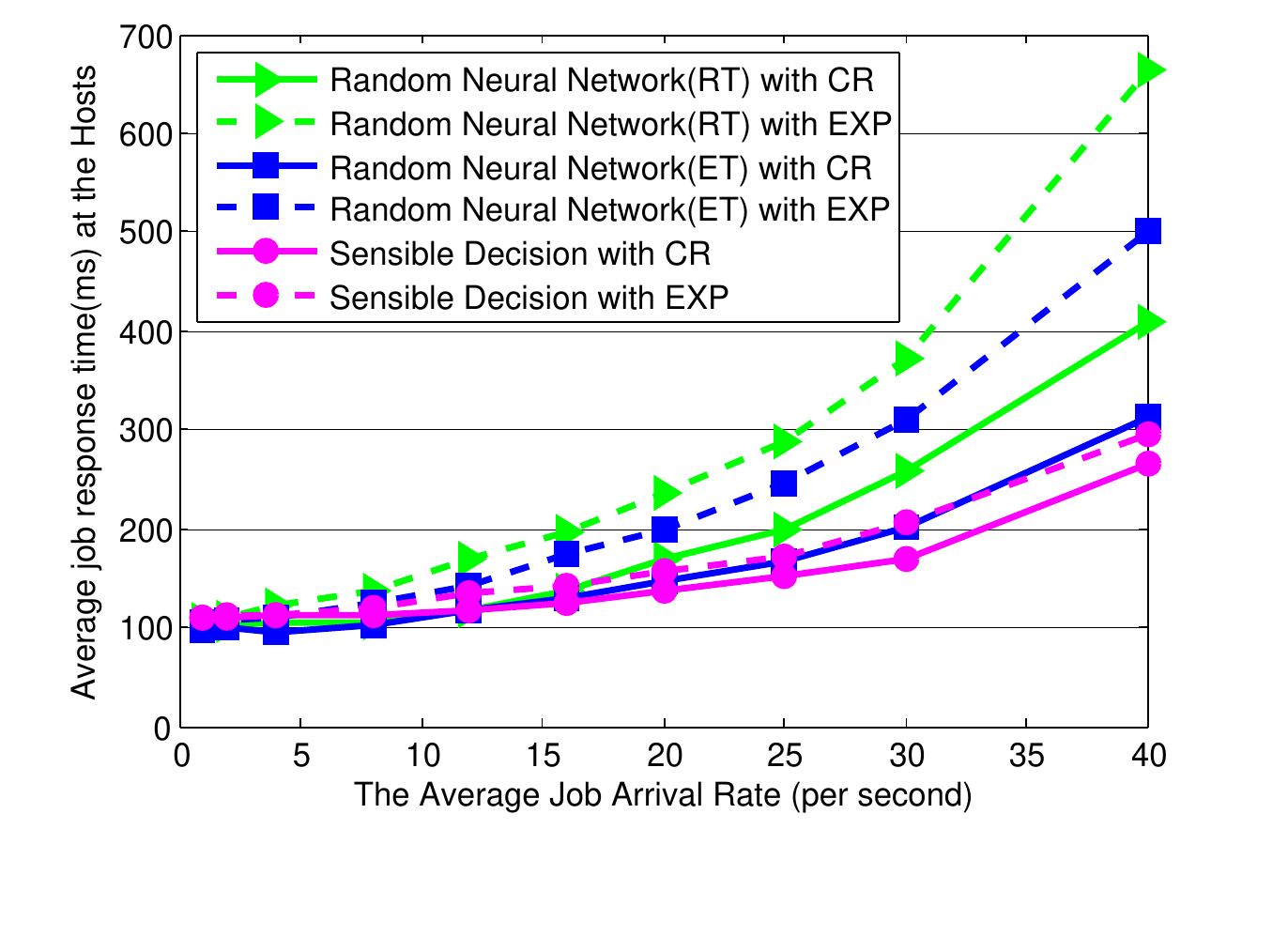}}
      \subfigure[\label{fig:cr_exp_ctrl}]{\includegraphics[scale=0.5,clip=true, trim=100 250 100 300,width=6.5cm,height=4.2cm]{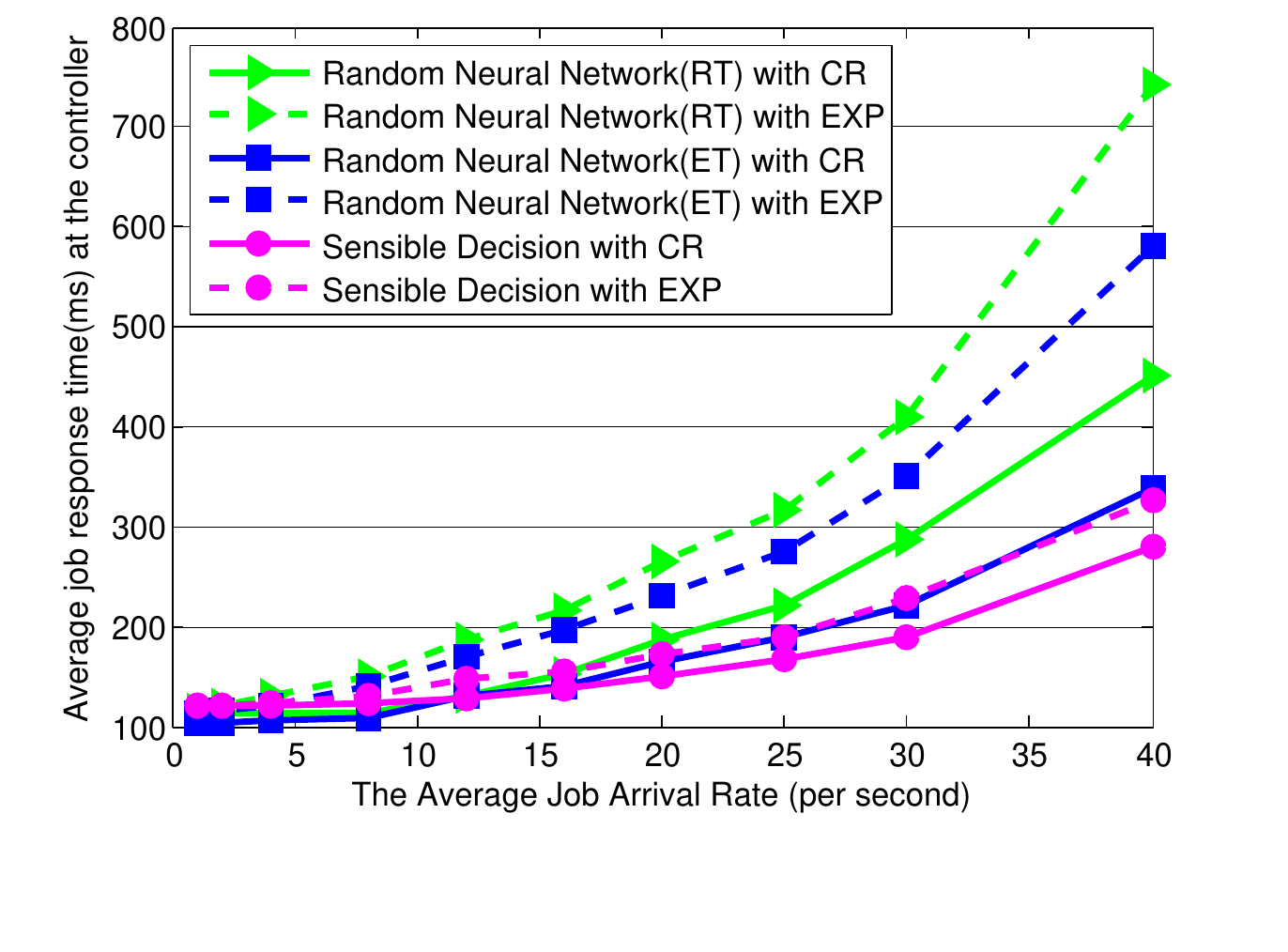}} 
			\subfigure[\label{fig:cr_exp_ctrl_zoomin}]{\includegraphics[scale=0.5,clip=true, trim=100 250 100 270,width=6.5cm,height=4.2cm]{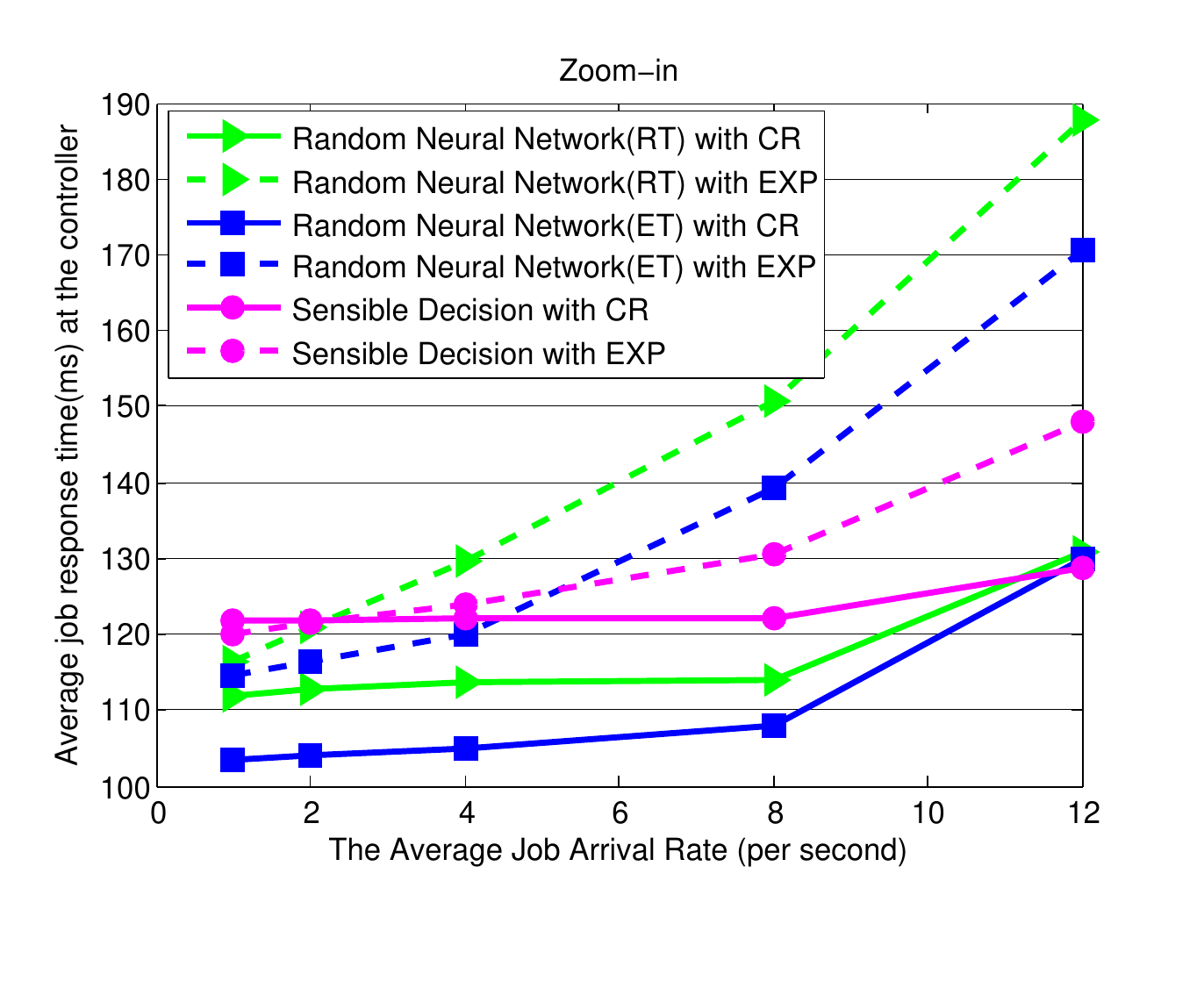}}
			\caption{The average job execution time, the average job response time at the hosts and the average job response time at the Controller in the six experiment scenarios versus the varied average job arrival rates }
			\label{fig:cr_exp}
     \end{figure}

\subsection{Comparison of the RNN and the Sensible Algorithm}
\label{performanceevaluationofrnnsensible}

We compared the two approaches with regard to task allocation based on the average job response time at the controller, the average job response time at the host and the average job execution time. The three metrics exhibit the same trend as shown in Figure~\ref{fig:cr_exp}.
 At low job arrival rates less than $8$/sec, the RNN with RL performs better as shown in Figure~\ref{fig:cr_exp_ctrl_zoomin}, and it is even  clearer with constant job arrival rates. 
However, as the average job arrival rates grows, the sensible decision algorithm outperforms the RNN, as in Figure~\ref{fig:cr_exp_ctrl}. Also the RNN algorithm with online measurement of the job execution time always performs better than the RNN with the metric of job response time. However, the sensible decision is always best under high job arrival rates, as shown in Figure~\ref{fig:cr_exp_ctrl} .
	

To explain these experimental results, we note that in these  we use CPU intensive jobs, and  {\em each of them} experience longer execution time than when they are executed separately due to the competition for the same physical resource, the CPU. Indeed, the hosts are multicore machines running Linux  with a multitasking capability so that multiple jobs will run together and interfere with each other 
as shown in Figure~\ref{fig:service_time_perjob}. It can be found that, for example, if four jobs running in parallel, the average execution/response time per job increases two times. That is to say, the fluctuation of the execution time that the jobs experienced under varied number of jobs in the system is quite significant. Since the RNN with RL will send the jobs to the best performing hosts, it will tend to overload them, contrary to the Sensible Algorithm which dispatches jobs probabilistically and therefore tends to spread the load in a better manner. 

\begin{figure}[ht]
		\centering
			\subfigure[\label{fig:job_alloc_RNN}]{\includegraphics[scale=0.6,clip=true, trim=130 300 130 300,width=6.5cm,height=4.5cm]{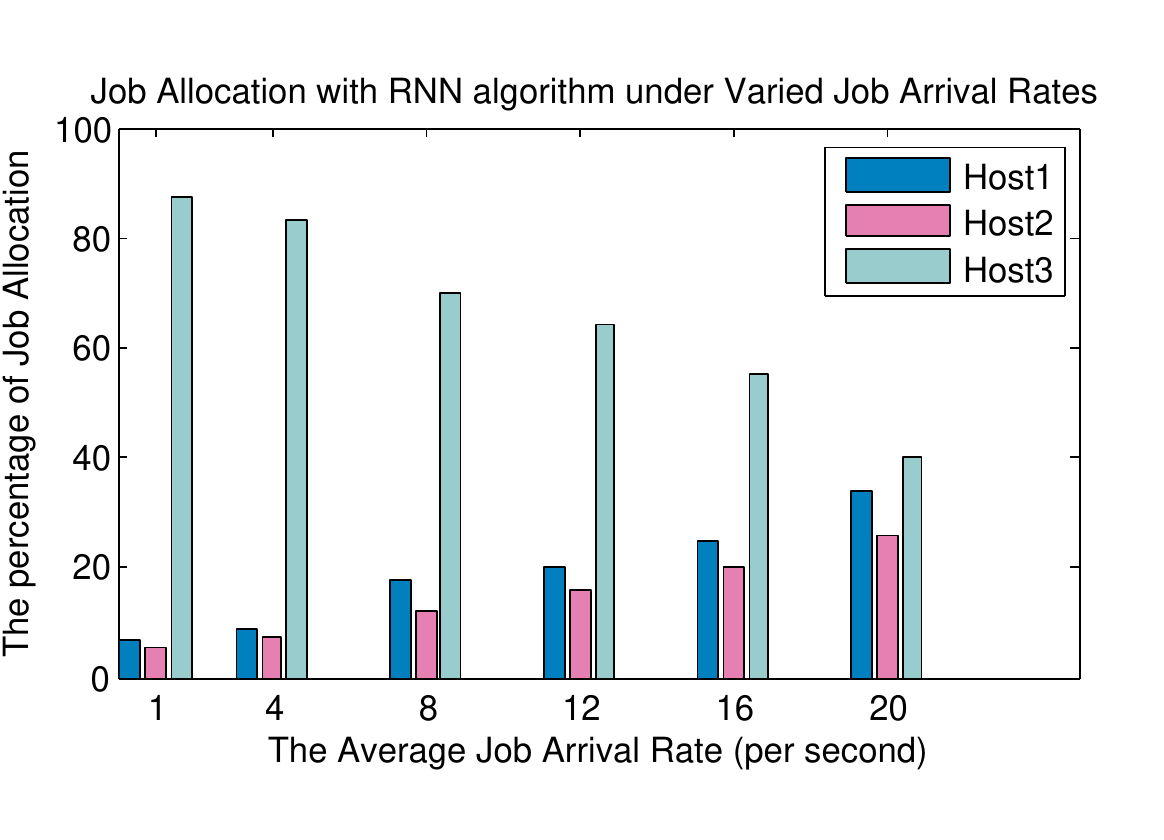}} 
		  \subfigure[\label{fig:job_alloc_Sens}]{\includegraphics[scale=0.6,clip=true, trim=130 300 120 300,width=6.5cm,height=4.5cm]{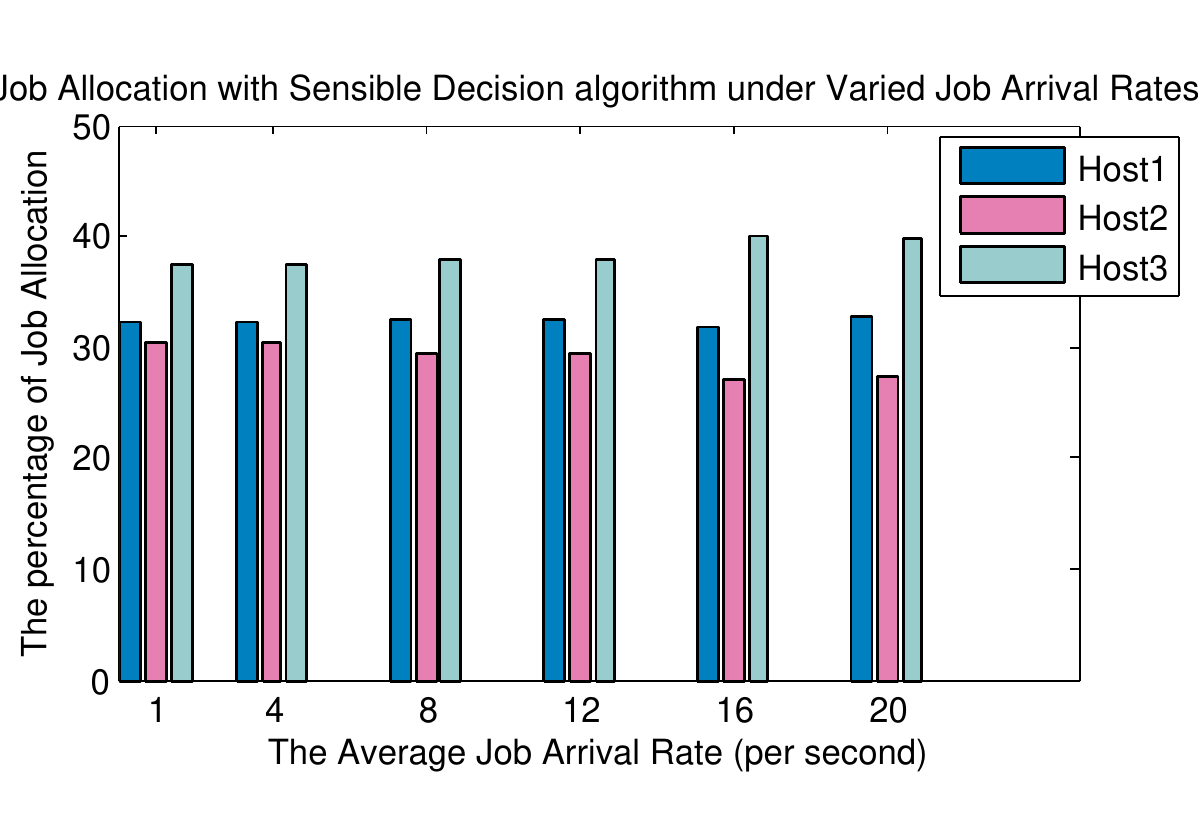}}
     \caption{The Proportion of job allocations to the three hosts with the RNN and the sensible decision algorithm for different arrival rates and Poisson job arrivals to TAP.}
   \end{figure}

When RNN used the job execution time as the QoS criterion, Figure~\ref{fig:job_alloc_RNN} shows that it dispatched the majority of jobs correctly to Host $3$  which provided the shortest service time. The other two hosts accommodated some jobs because the RNN algorithm was programmed to make 10\% of its decisions at random with equal probability. Here, the sensible decision algorithm performed worse because it makes job allocation decision with a probability that is inversely proportional to the job response time/execution time, instead of exactly following the best QoS as the RNN. As shown in Figure~\ref{fig:job_alloc_Sens}, the proportion of the jobs allocated with the sensible decision algorithm coincides with the proportion of the respective speeds of the three hosts.

On the other hand, the sensible decision algorithm benefits from the fact that it does not overload the ``best'' hosts as shown in Figure~\ref{fig:cr_exp_ctrl} where the jobs may sometimes arrive to a host at rate that is higher than the average processing rate.  In Figure~\ref{fig:cr_exp} we also see that the RNN based algorithm, that uses the job execution time measured at the hosts as the QoS goal,  outperforms the RNN with online measurement of the job response time,  because the actual job execution can be a more accurate predictor of overall performance when the communication times between the hosts and the TAP
fluctuate significantly. However at high job arrival rates,  the sensible decision algorithm again performed better. 

\subsection{Comparison with the Model Based and Static Allocation Schemes}
\label{performanceevaluationofallschemes}

\begin{figure}[ht]
	
		\centering
			\subfigure[\label{fig:expet}]{\includegraphics[scale=0.5,clip=true, trim=100 250 130 300,width=6.5cm,height=4.5cm]{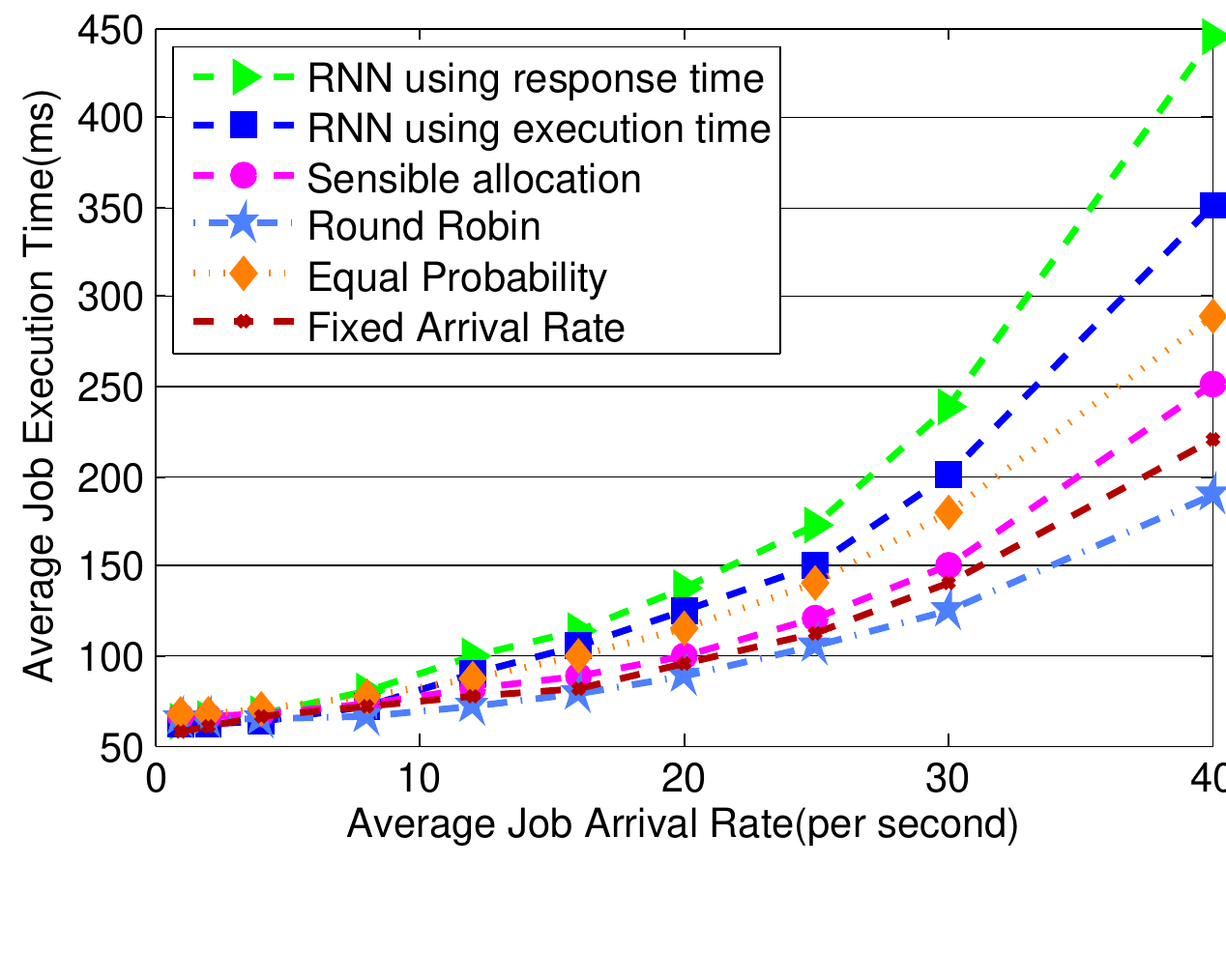}} 
		  \subfigure[\label{fig:expetzoomin}]{\includegraphics[scale=0.5,clip=true, trim=100 250 120 300,width=6.5cm,height=4.5cm]{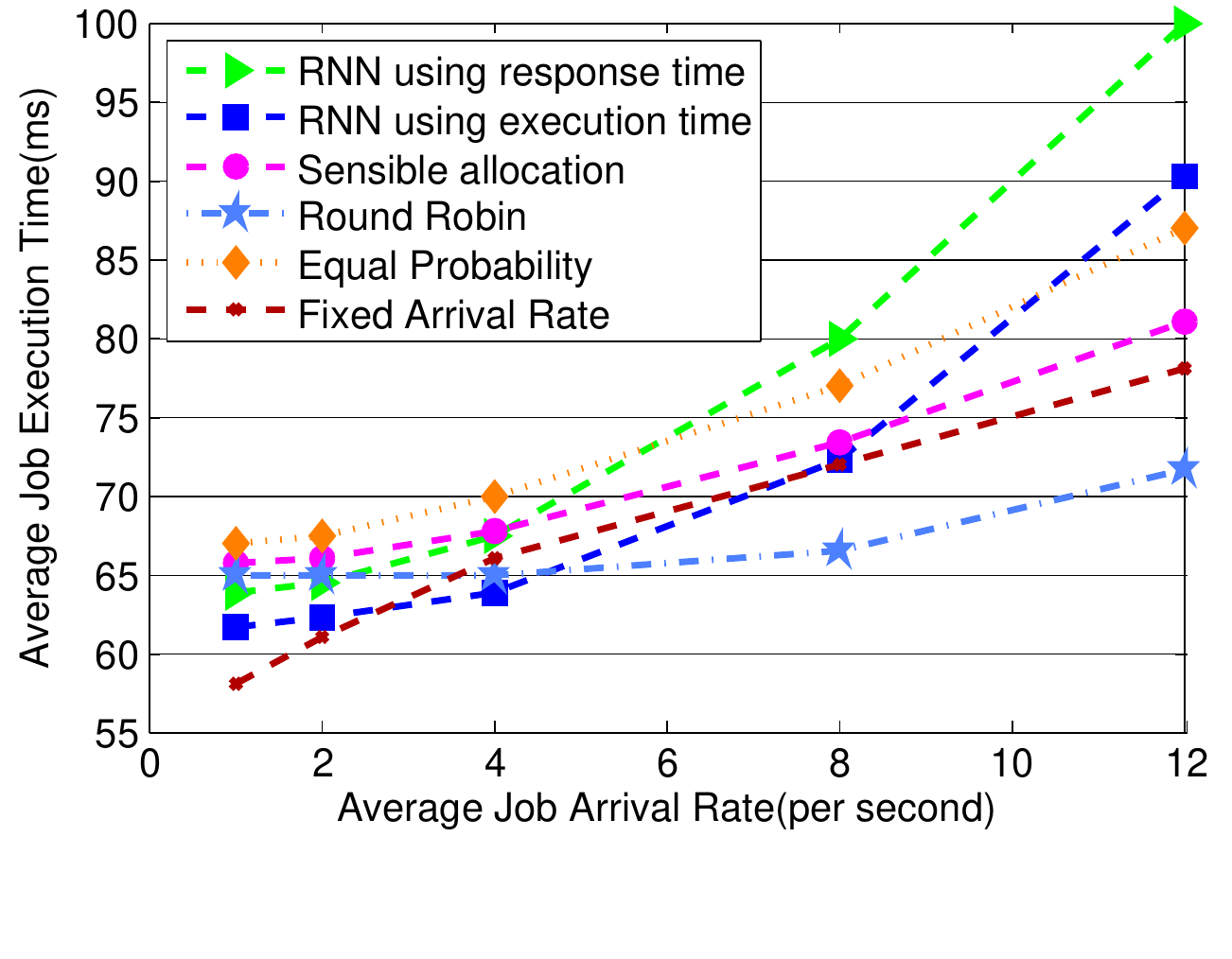}}
     \caption{The average execution time experienced under varied job arrival rates and different task allocation schemes when the three hosts have similar performance.}
		\label{fig:etwithallalgo}
   \end{figure}

Figure~\ref{fig:etwithallalgo} shows  the average job execution time obtained with the RNN and the Sensible Algorithm, in comparison with the model based
scheme, as well as the Round Robin and Equally Probable allocation. The model based scheme performed better than the RNN when the job arrival rate was low, and better than the Sensible Algorithm 
at high arrival rates. 
However, the model based scheme can be viewed as an ``ideal benchmark'' since it relies on full information: it assumes knowledge of the arrival rate, it supposes that arrivals are Poisson,
and it assumes knowledge of the job service rates at each host, while the RNN based scheme just observes the most recent measurement of the goal function.

As expected the equally probable allocation scheme performed worse. In this case where all servers are roughly equivalent in speed,  Round Robin always outperformed the Sensible Algorithm, because it distributes work in a manner that does not overload any of the servers. These results are summarized in  Figure ~\ref{fig:expet}. However the observed results change when the hosts have distinct performance characteristics as shown below.

		\begin{figure}[ht]
		
		\centering
			\subfigure[\label{fig:expetstressed}]{\includegraphics[scale=0.5,clip=true, trim=100 250 130 304,width=6.5cm,height=4.5cm]{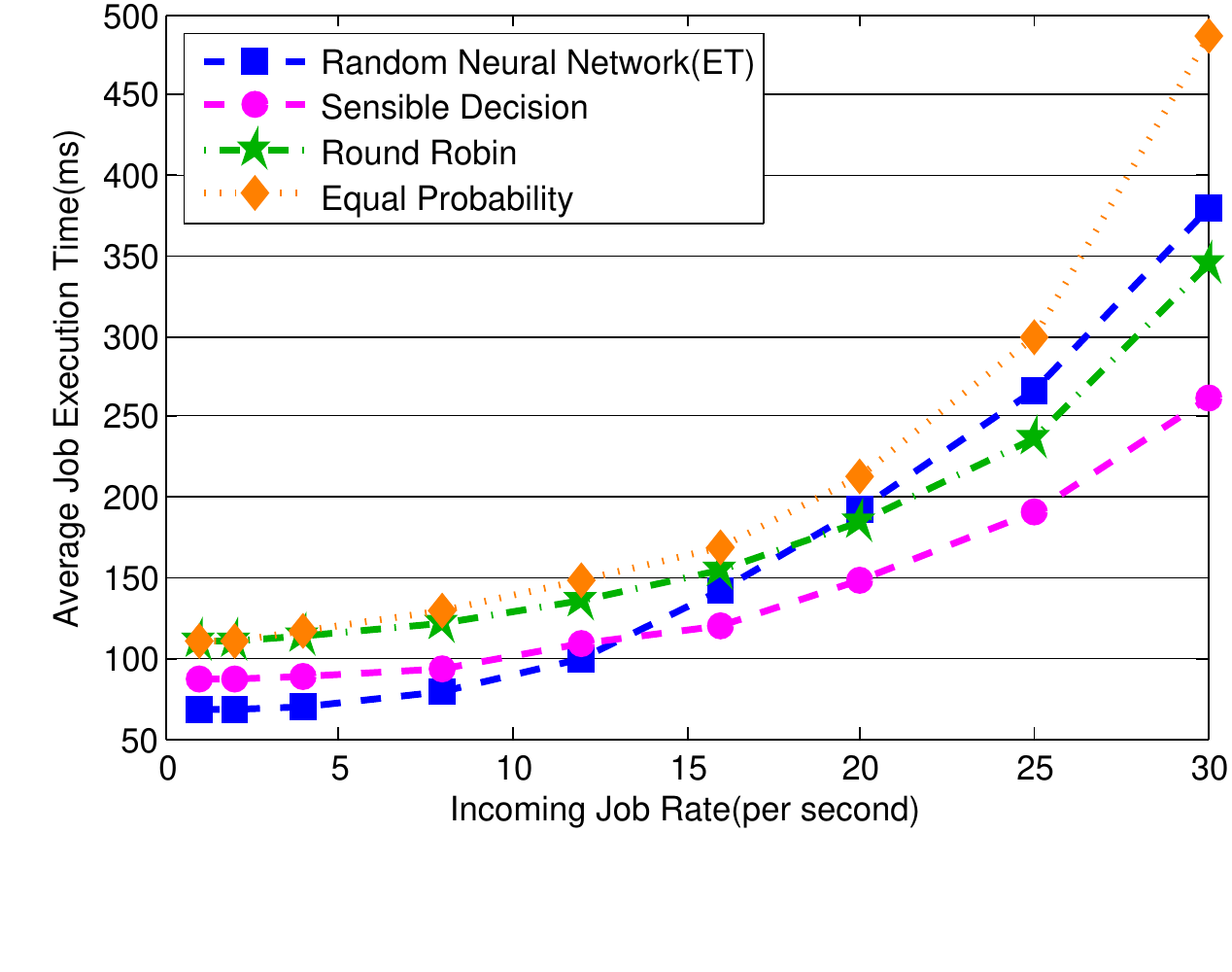}} 
		  \subfigure[\label{fig:expetstressedzoomin}]{\includegraphics[scale=0.5,clip=true, trim=100 250 120 300,width=6.5cm,height=4.5cm]{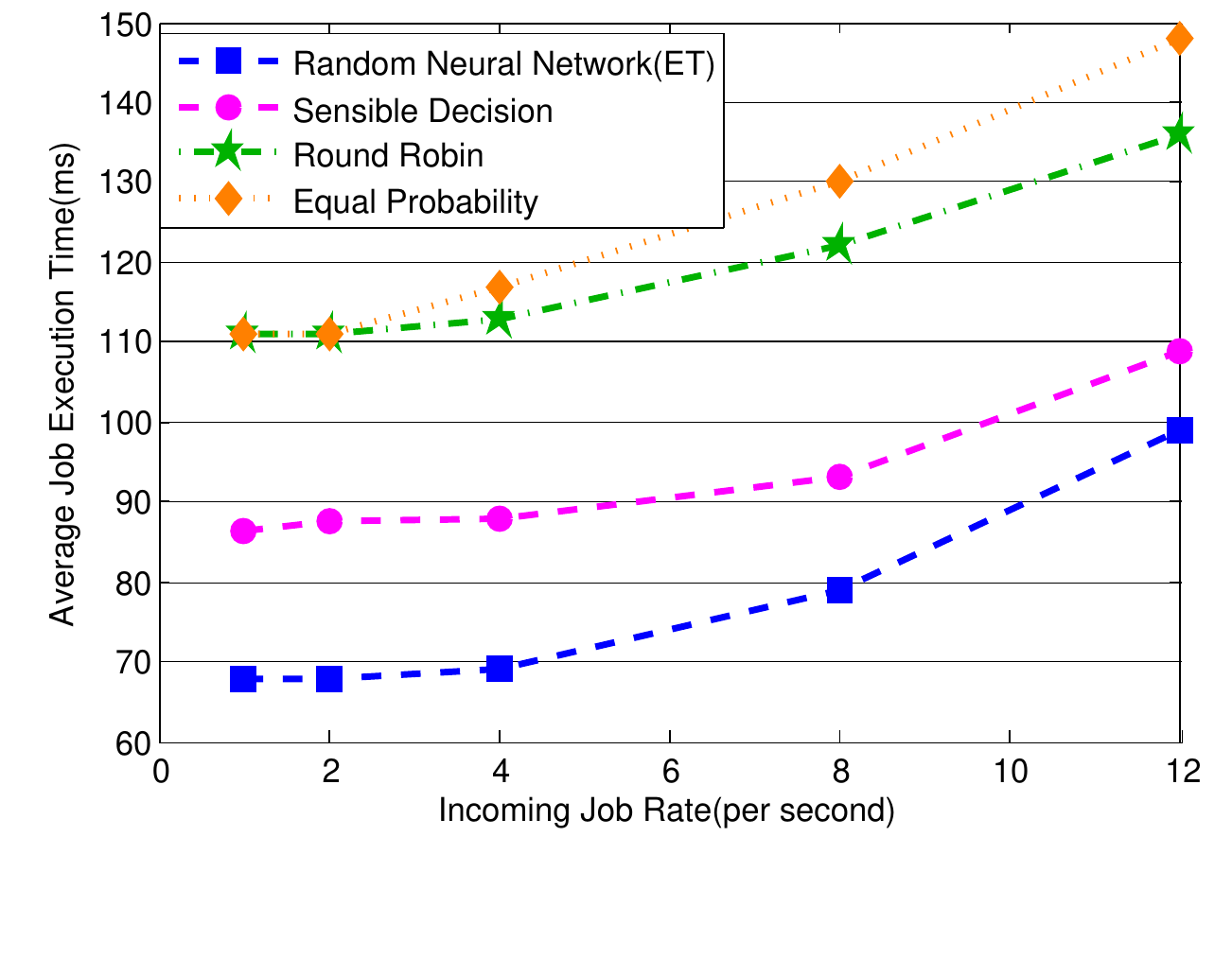}}
     \caption{Average execution time experienced in a cluster composed of hosts with non-uniform processing capacities.}
		\label{fig:et_stressed}
   \end{figure}
	
\subsection{Performance Measurements when Hosts have Distinct Processing Rates}
\label{performanceevaluationununiform}

As a last step, we evaluate the algorithms that we have considered, in a situation where each hosts provides significantly different performance. Since the hosts we have for our experiments are quite similar, we 
introduced a background load on each host which runs constantly and independently of the tasks that TAP allocates to the hosts. This is in fact a realistic situation since in a Cloud,
multiple sources of tasks may share the same set of hosts without knowing what their precise workload may be, except for external observations of their performance.

Thus we were able to emulate as set of three Hosts $1,~2,~3$ with relative processing speeds of $2:4:1$. The results of these experiments are
 summarized in Figure~\ref{fig:et_stressed}. 
 We see that TAP with both the RNN and the Sensible Algorithm benefits from the ability of these two schemes to measure the performance differences between the hosts, and dispatch jobs to the hosts which offer a better  performance, whereas the two static allocation schemes (Round Robin and the allocation of tasks with equal probabilities) lead to worse performance as a whole. 
 
 The performance of the RNN-base scheme clearly stands out among the others as shown in Figure~\ref{fig:expetstressedzoomin}, confirming that a system such as TAP equipped with the RNN 
 can provide a very useful fine-grained QoS-aware task allocation algorithm.

\section{Conclusions and Future Work} \label{conclusions}

In this paper we have presented TAP, a task allocation platform which can incorporate a variety of different algorithms to dispatch jobs to hosts in the Cloud.
TAP can exploit both simple static allocations schemes (such as the Round Robin), as well as measurement driven adaptive on-line algorithms such as the 
RNN and the Sensible Algorithms that bring intelligence to bear from observations and make judicious allocation decisions. 
We conducted numerous experiments with a CPU intensive workload to evaluate both static and adaptive allocation schemes in two different hosting environments: one composed of hosts with very similar processing speeds, and another one with hosts having different speeds due to distinct background loads at each host. 

Experiments showed that when the hosts are quite distinct, the RNN based algorithm with Reinforcement-Learning offered a fine-grained QoS-aware task allocation algorithm which can make accurate decisions provided that the online measurements are regularly updated. We found that the Sensible Algorithm offers a robust QoS-aware scheme with the potential to perform better under loads. The fixed arrival
rate scheme, with full information of arrival rates and service rates,  outperformed both the RNN and ``sensible'' approach  due to the fact that it employs the solution of an analytical model 
that allows one to minimize job response time under known mathematical assumptions which may actually not known or valid in practice: it is thus useful as a benchmark but cannot be recommended in practical situations. Round Robin is a simple algorithm, which is effective when the processing rates and loads at each of the hosts are very similar. 
	
In future work we will investigate the use of more sophisticated mathematical models such as diffusions approximations \cite{Diffusion} to build a model driven allocation algorithm that exploits on-line measurements of the arrival and service statistics at each of the hosts in order to estimate the task allocation probabilities. Although we expect that such an approach will have its limits due to the increase of the
amount of data that it will need, we also think that it may offer a better benchmark for the comparison of various allocation methods. We would also like to study the Cloud system we have described when a given set of hosts
is used by multiple TAP systems with heterogenous input streams (such as Web services, mobile services and compute intensive applications) to see which schemes can offer the most robust and resilient
allocation schemes in the presence of competing and diverse workloads. Another direction we wish to undertake is the study of the robustness of allocation schemes
for Cloud services in the presence of attacks \cite{DDoS} designed to disrupt normal operations.




\bibliographystyle{IEEEtran}
\bibliography{references}

\begin{thebibliography}{10}
\providecommand{\url}[1]{#1}
\csname url@samestyle\endcsname
\providecommand{\newblock}{\relax}
\providecommand{\bibinfo}[2]{#2}
\providecommand{\BIBentrySTDinterwordspacing}{\spaceskip=0pt\relax}
\providecommand{\BIBentryALTinterwordstretchfactor}{4}
\providecommand{\BIBentryALTinterwordspacing}{\spaceskip=\fontdimen2\font plus
\BIBentryALTinterwordstretchfactor\fontdimen3\font minus
  \fontdimen4\font\relax}
\providecommand{\BIBforeignlanguage}[2]{{%
\expandafter\ifx\csname l@#1\endcsname\relax
\typeout{** WARNING: IEEEtran.bst: No hyphenation pattern has been}%
\typeout{** loaded for the language `#1'. Using the pattern for}%
\typeout{** the default language instead.}%
\else
\language=\csname l@#1\endcsname
\fi
#2}}
\providecommand{\BIBdecl}{\relax}
\BIBdecl

\bibitem{mell2009}
P.~Mell and T.~Grance, ``The nist definition of cloud computing,'' \emph{NIST
  Special Publication 800-145}, Aug 2009.

\bibitem{buyya2013}
R.~Buyya, ``Introduction to the ieee transactions on cloud computing,''
  \emph{Cloud Computing, IEEE Transactions on}, vol.~1, no.~1, pp. 3--21, Jan
  2013.

\bibitem{Delimitrou2013}
\BIBentryALTinterwordspacing
C.~Delimitrou and C.~Kozyrakis, ``Qos-aware scheduling in heterogeneous
  datacenters with paragon,'' \emph{ACM Trans. Comput. Syst.}, vol.~31, no.~4,
  pp. 12:1--12:34, Dec. 2013. [Online]. Available:
  \url{http://doi.acm.org/10.1145/2556583}
\BIBentrySTDinterwordspacing

\bibitem{Padala2007}
\BIBentryALTinterwordspacing
P.~Pradeep, K.~G. Shin, X.~Zhu, M.~Uysal, Z.~Wang, S.~Singhal, A.~Merchant, and
  K.~Salem, ``Adaptive control of virtualized resources in utility computing
  environments,'' in \emph{Proceedings of the 2Nd ACM SIGOPS/EuroSys European
  Conference on Computer Systems 2007}, ser. EuroSys '07.\hskip 1em plus 0.5em
  minus 0.4em\relax New York, NY, USA: ACM, 2007, pp. 289--302. [Online].
  Available: \url{http://doi.acm.org/10.1145/1272996.1273026}
\BIBentrySTDinterwordspacing

\bibitem{berl2010energy}
A.~Berl, E.~Gelenbe, M.~Di~Girolamo, G.~Giuliani, H.~De~Meer, M.~Q. Dang, and
  K.~Pentikousis, ``Energy-efficient cloud computing,'' \emph{The Computer
  Journal}, vol.~53, no.~7, pp. 1045--1051, 2010.

\bibitem{GelenbeL13}
\BIBentryALTinterwordspacing
E.~Gelenbe and R.~Lent, ``Energy-qos trade-offs in mobile service selection,''
  \emph{Future Internet}, vol.~5, no.~2, pp. 128--139, 2013. [Online].
  Available: \url{http://dx.doi.org/10.3390/fi5020128}
\BIBentrySTDinterwordspacing

\bibitem{thai13}
------, ``Optimising server energy consumption and response time,''
  \emph{Theoretical and Applied Informatics}, no.~4, pp. 257--270, Jan 2013.

\bibitem{Jianfeng2013}
J.~Zhan, L.~Wang, X.~Li, W.~Shi, C.~Weng, W.~Zhang, and X.~Zang, ``Cost-aware
  cooperative resource provisioning for heterogeneous workloads in data
  centers,'' \emph{Computers, IEEE Transactions on}, vol.~62, no.~11, pp.
  2155--2168, Nov 2013.

\bibitem{Iosup2011}
A.~Iosup, S.~Ostermann, M.~Yigitbasi, R.~Prodan, T.~Fahringer, and D.~H.~J.
  Epema, ``Performance analysis of cloud computing services for many-tasks
  scientific computing,'' \emph{Parallel and Distributed Systems, IEEE
  Transactions on}, vol.~22, no.~6, pp. 931--945, June 2011.

\bibitem{Zhuravlev}
\BIBentryALTinterwordspacing
S.~Zhuravlev, S.~Blagodurov, and A.~Fedorova, ``Addressing shared resource
  contention in multicore processors via scheduling,'' \emph{SIGPLAN Not.},
  vol.~45, no.~3, pp. 129--142, Mar. 2010. [Online]. Available:
  \url{http://doi.acm.org/10.1145/1735971.1736036}
\BIBentrySTDinterwordspacing

\bibitem{tantawi1985optimal}
A.~N. Tantawi and D.~Towsley, ``Optimal static load balancing in distributed
  computer systems,'' \emph{Journal of the ACM (JACM)}, vol.~32, no.~2, pp.
  445--465, 1985.

\bibitem{kim1992algorithm}
C.~Kim and H.~Kameda, ``An algorithm for optimal static load balancing in
  distributed computer systems,'' \emph{IEEE Transactions on Computers},
  vol.~41, no.~3, pp. 381--384, 1992.

\bibitem{kameda2011optimal}
H.~Kameda, J.~Li, C.~Kim, and Y.~Zhang, \emph{Optimal Load Balancing in
  Distributed Computer Systems}.\hskip 1em plus 0.5em minus 0.4em\relax
  Springer, 2011.

\bibitem{wolff2001dynamic}
J.~J. Wolff, ``Dynamic load balancing of a network of client and server
  computers,'' Feb.~6, 2001, u. S. Patent 6,185,601.

\bibitem{rimal2009taxonomy}
B.~P. Rimal, E.~Choi, and I.~Lumb, ``A taxonomy and survey of cloud computing
  systems,'' in \emph{INC, IMS and IDC, 2009. NCM'09. Fifth International Joint
  Conference on}.\hskip 1em plus 0.5em minus 0.4em\relax Ieee, 2009, pp.
  44--51.

\bibitem{zhang2010load}
Z.~Zhang and X.~Zhang, ``A load balancing mechanism based on ant colony and
  complex network theory in open cloud computing federation,'' in
  \emph{Industrial Mechatronics and Automation (ICIMA), 2010 2nd International
  Conference on}, vol.~2.\hskip 1em plus 0.5em minus 0.4em\relax IEEE, 2010,
  pp. 240--243.

\bibitem{tian2011dynamic}
W.~Tian, Y.~Zhao, Y.~Zhong, M.~Xu, and C.~Jing, ``A dynamic and integrated
  load-balancing scheduling algorithm for cloud datacenters,'' in \emph{Cloud
  Computing and Intelligence Systems (CCIS), 2011 IEEE International Conference
  on}.\hskip 1em plus 0.5em minus 0.4em\relax IEEE, 2011, pp. 311--315.

\bibitem{Xiaomin2011}
X.~Zhu, X.~Qin, and M.~Qiu, ``Qos-aware fault-tolerant scheduling for real-time
  tasks on heterogeneous clusters,'' \emph{Computers, IEEE Transactions on},
  vol.~60, no.~6, pp. 800--812, June 2011.

\bibitem{Topcuouglu2002}
\BIBentryALTinterwordspacing
H.~Topcuouglu, S.~Hariri, and M.~you Wu, ``Performance-effective and
  low-complexity task scheduling for heterogeneous computing,'' \emph{IEEE
  Trans. Parallel Distrib. Syst.}, vol.~13, no.~3, pp. 260--274, Mar. 2002.
  [Online]. Available: \url{http://dx.doi.org/10.1109/71.993206}
\BIBentrySTDinterwordspacing

\bibitem{Sih1993}
G.~Sih and E.~Lee, ``A compile-time scheduling heuristic for
  interconnection-constrained heterogeneous processor architectures,''
  \emph{Parallel and Distributed Systems, IEEE Transactions on}, vol.~4, no.~2,
  pp. 175--187, Feb 1993.

\bibitem{Kwok1996}
Y.-K. Kwok and I.~Ahmad, ``Dynamic critical-path scheduling: an effective
  technique for allocating task graphs to multiprocessors,'' \emph{Parallel and
  Distributed Systems, IEEE Transactions on}, vol.~7, no.~5, pp. 506--521, May
  1996.

\bibitem{Hou1994}
E.~Hou, N.~Ansari, and H.~Ren, ``A genetic algorithm for multiprocessor
  scheduling,'' \emph{Parallel and Distributed Systems, IEEE Transactions on},
  vol.~5, no.~2, pp. 113--120, Feb 1994.

\bibitem{WeiNeng2009}
W.~Chen and J.~Zhang, ``An ant colony optimization approach to a grid workflow
  scheduling problem with various qos requirements,'' \emph{Systems, Man, and
  Cybernetics, Part C: Applications and Reviews, IEEE Transactions on},
  vol.~39, no.~1, pp. 29--43, Jan 2009.

\bibitem{Pandey2010}
S.~Pandey, W.~Linlin, S.~Guru, and R.~Buyya, ``A particle swarm
  optimization-based heuristic for scheduling workflow applications in cloud
  computing environments,'' in \emph{Advanced Information Networking and
  Applications (AINA), 2010 24th IEEE International Conference on}, April 2010,
  pp. 400--407.

\bibitem{DBLP:journals/neco/GelenbeF99}
\BIBentryALTinterwordspacing
E.~Gelenbe and J.~Fourneau, ``Random neural networks with multiple classes of
  signals,'' \emph{Neural Computation}, vol.~11, no.~4, pp. 953--963, 1999.
  [Online]. Available: \url{http://dx.doi.org/10.1162/089976699300016520}
\BIBentrySTDinterwordspacing

\bibitem{Gelenbe2010_nearOptAssign}
\BIBentryALTinterwordspacing
E.~Gelenbe, S.~Timotheou, and D.~Nicholson, ``Fast distributed near-optimum
  assignment of assets to tasks,'' \emph{The Computer Journal}, vol.~53, no.~9,
  pp. 1360--1369, Nov. 2010. [Online]. Available:
  \url{http://dx.doi.org/10.1093/comjnl/bxq010}
\BIBentrySTDinterwordspacing

\bibitem{Zaman2011}
S.~Zaman and D.~Grosu, ``A combinatorial auction-based dynamic vm provisioning
  and allocation in clouds,'' in \emph{Cloud Computing Technology and Science
  (CloudCom), 2011 IEEE Third International Conference on}, Nov 2011, pp.
  107--114.

\bibitem{Lin2011}
C.~Lin and S.~Lu, ``Scheduling scientific workflows elastically for cloud
  computing,'' in \emph{Cloud Computing (CLOUD), 2011 IEEE International
  Conference on}, July 2011, pp. 746--747.

\bibitem{Moreno2014}
I.~S. Moreno, P.~Garraghan, P.~Townend, and J.~Xu, ``Analysis, modeling and
  simulation of workload patterns in a large-scale utility cloud,'' \emph{Cloud
  Computing, IEEE Transactions on}, vol.~PP, no.~99, pp. 1--1, 2014.

\bibitem{Palanisamy2014}
B.~Palanisamy, A.~Singh, and L.~Liu, ``Cost-effective resource provisioning for
  mapreduce in a cloud,'' \emph{Parallel and Distributed Systems, IEEE
  Transactions on}, vol.~PP, no.~99, pp. 1--1, 2014.

\bibitem{Zhuravlev2010}
\BIBentryALTinterwordspacing
S.~Zhuravlev, S.~Blagodurov, and A.~Fedorova, ``Addressing shared resource
  contention in multicore processors via scheduling,'' \emph{SIGPLAN Not.},
  vol.~45, no.~3, pp. 129--142, Mar. 2010. [Online]. Available:
  \url{http://doi.acm.org/10.1145/1735971.1736036}
\BIBentrySTDinterwordspacing

\bibitem{Bhatti1999}
N.~Bhatti and R.~Friedrich, ``Web server support for tiered services,''
  \emph{Network, IEEE}, vol.~13, no.~5, pp. 64--71, Sep 1999.

\bibitem{Ying2013}
Y.~Song, Y.~Sun, and W.~Shi, ``A two-tiered on-demand resource allocation
  mechanism for vm-based data centers,'' \emph{Services Computing, IEEE
  Transactions on}, vol.~6, no.~1, pp. 116--129, First 2013.

\bibitem{douratsos}
\BIBentryALTinterwordspacing
E.~Gelenbe, R.~Lent, and M.~Douratsos, ``Choosing a local or remote cloud,'' in
  \emph{Second Symposium on Network Cloud Computing and Applications, {NCCA}
  2012, London, United Kingdom, December 3-4, 2012}.\hskip 1em plus 0.5em minus
  0.4em\relax IEEE Computer Society, 2012, pp. 25--30. [Online]. Available:
  \url{http://doi.ieeecomputersociety.org/10.1109/NCCA.2012.16}
\BIBentrySTDinterwordspacing

\bibitem{gelenbe2012trade}
E.~Gelenbe and R.~Lent, ``Trade-offs between energy and quality of service,''
  in \emph{Sustainable Internet and ICT for Sustainability (SustainIT),
  2012}.\hskip 1em plus 0.5em minus 0.4em\relax IEEE, 2012, pp. 1--5.

\bibitem{Zhang2014}
Q.~Zhang, M.~Zhani, R.~Boutaba, and J.~Hellerstein, ``Dynamic
  heterogeneity-aware resource provisioning in the cloud,'' \emph{Cloud
  Computing, IEEE Transactions on}, vol.~2, no.~1, pp. 14--28, March 2014.

\bibitem{dillon2010cloud}
T.~Dillon, C.~Wu, and E.~Chang, ``Cloud computing: issues and challenges,'' in
  \emph{Advanced Information Networking and Applications (AINA), 2010 24th IEEE
  International Conference on}.\hskip 1em plus 0.5em minus 0.4em\relax Ieee,
  2010, pp. 27--33.

\bibitem{SAN04}
E.~Gelenbe, R.~Lent, and A.~Nunez, ``Self-aware networks and {QoS},''
  \emph{Proceedings of the {IEEE}}, vol.~92, no.~9, pp. 1478--1489, Sep. 2004.

\bibitem{DBLP:journals/neco/GelenbeT08}
\BIBentryALTinterwordspacing
E.~Gelenbe and S.~Timotheou, ``Random neural networks with synchronized
  interactions,'' \emph{Neural Computation}, vol.~20, no.~9, pp. 2308--2324,
  2008. [Online]. Available:
  \url{http://dx.doi.org/10.1162/neco.2008.04-07-509}
\BIBentrySTDinterwordspacing

\bibitem{gelenbe2003sensible}
E.~Gelenbe, ``Sensible decisions based on qos,'' \emph{Computational management
  science}, vol.~1, no.~1, pp. 1--14, 2003.

\bibitem{TAI}
E.~Gelenbe, Z.~Xu, and E.~Seref, ``Cognitive packet networks,'' in
  \emph{Proceedings of the 11th International Conference on Tools with
  Artificial Intelligence}, Nov. 1999, pp. 47--54.

\bibitem{network1}
E.~Gelenbe, ``Steps toward self-aware networks,'' \emph{Communications ACM},
  vol.~52, no.~7, pp. 66--75, July 2009.

\bibitem{Sutton}
R.~S. Sutton and A.~G. Barto, \emph{Reinforcement Learning: An
  Introduction}.\hskip 1em plus 0.5em minus 0.4em\relax MIT Press, 1998.

\bibitem{Gelenbe-Mitrani}
E.~Gelenbe and I.~Mitrani, \emph{Analysis and Synthesis of Computer
  Systems}.\hskip 1em plus 0.5em minus 0.4em\relax World Scientific, Imperial
  College Press, 2010.

\bibitem{Diffusion}
\BIBentryALTinterwordspacing
E.~Gelenbe, ``Probabilistic models of computer systems,'' \emph{Acta
  Informatica}, vol.~12, pp. 285--303, 1979. [Online]. Available:
  \url{http://dx.doi.org/10.1007/BF00268317}
\BIBentrySTDinterwordspacing

\bibitem{RNN}
E.~Gelenbe and J.~Fourneau, ``Random neural networks with multiple classes of
  signals,'' \emph{Neural Computation}, vol.~11, no.~4, pp. 953--963, 1999.

\bibitem{Gelenbe1990_RNN}
E.~Gelenbe, ``Stability of the random neural network model,'' \emph{Neural
  Computation}, vol.~2, no.~2, pp. 239--247, 1990.

\bibitem{Gelenbe01112008}
E.~Gelenbe and S.~Timotheou, ``Synchronized interactions in spiked neuronal
  networks,'' \emph{The Computer Journal}, vol.~51, no.~6, pp. 723--730, 2008.

\bibitem{DDoS}
E.~Gelenbe and G.~Loukas, ``A self-aware approach to denial of service
  defence,'' \emph{Computer Networks}, vol.~51, no.~5, pp. 1299--1314, April
  2007.

\end{thebibliography}

\end{document}